\documentclass[useAMS,usenatbib]{mnras}
\usepackage{amsmath,amsfonts}
\usepackage{amssymb}
\usepackage{graphicx}
\usepackage{nth}
\usepackage{subfig}
\usepackage{url}
\usepackage{multirow}
\usepackage{enumerate}
\usepackage[skip=10pt]{caption}
\newcommand{\ions}[1]{\,{\rm {\scriptstyle #1}}}

\title{Enabling Radiative Transfer on AMR grids in CRASH}
\author[Hariharan N. et al.]{N. Hariharan$^{1,4}$\thanks{Email:nty@mpa-garching.mpg.de}, L. Graziani$^{1,5}$, B. Ciardi$^{1}$, F.
Miniati$^{2}$, H.-J. Bungartz$^{3}$\\
$^{1}$Max-Planck-Instit\"{u}t f\"{u}r Astrophysik, Karl-Schwarzschild-Strasse 1,
85748 Garching, Germany\\
$^{2}$Department of Physics, ETH-Z\"{u}rich, Wolfgang-Pauli-Strasse 27, CH-8093,
Z\"{u}rich, Switzerland\\
$^{3}$Instit\"{u}t f\"{u}r Informatik, TU M\"{u}nchen, D-80290, M\"{u}nchen,
Germany\\
$^{4}$Intel Technology India Private Limited, \#136 Old Airport Road, Bangalore-560008, India\\
$^{5}$INAF Osservatorio Astronomico di Roma, Via Frascati 33, 00040, Monte Porzio Catone (RM), Italy}

\begin{document}

\raggedbottom

\date{Accepted 1988 December 15. Received 1988 December 14; in original form
1988 October 11}

\pagerange{\pageref{firstpage}--\pageref{lastpage}} \pubyear{2015}

\maketitle

\label{firstpage}

\begin{abstract}


We introduce \texttt{CRASH-AMR}, a new version of the cosmological Radiative
Transfer (RT) code \texttt{CRASH}, enabled to use refined grids. This new
feature allows us to attain higher resolution in our RT simulations and thus to
describe more accurately ionisation and temperature patterns in high density
regions. We have tested \texttt{CRASH-AMR} by simulating the evolution of an
ionised region produced by a single source embedded in gas at constant density,
as well as by a more realistic configuration of multiple sources in an
inhomogeneous density field. While we find an excellent agreement with the
previous version of \texttt{CRASH} when the AMR feature is disabled, showing
that no numerical artifact has been introduced in \texttt{CRASH-AMR}, when
additional refinement levels are used the code can simulate more accurately the
physics of ionised gas in high density regions. 
This result has been attained at no computational loss, as RT simulations on AMR
grids with maximum resolution equivalent to that of a uniform cartesian grid can
be run with a gain of up to 60\% in computational time.
\end{abstract}

\section{Introduction}

The observed universe shows a large variation in structures as we move along
different scales. Independent observations of the distribution of faint radio
sources, optically selected galaxies and the X-ray background show that our
universe is homogeneous on a scale larger than 200~Mpc (see e.g.
\citealt{Wu1999} for a review), while at smaller scales it appears inhomogeneous
due to the presence of a large number of structures: for example, galaxy
clusters and galaxies at tens of Mpc, or stellar clusters and molecular clouds
at parsec scales. In the $\Lambda$CDM concordance model of our universe, the
presence of primordial density perturbations led to the gravitational collapse
and cooling of gas in pre-existing dark-matter halos, and to the subsequent
formation of radiating sources like stars and quasars (e.g. \citealt{Mo2010}).
The chemical, mechanical and radiative feedback by these sources on their
surroundings induces a complex interplay between the galaxy formation process
and the evolution of the intergalactic medium (IGM; \citealt{Ciardietal2005,
Dave2005, Barkana2007, Meiksin2009}). Among these feedback effects, a prominent
place is occupied by the IGM reionisation process, which denotes the transition
from a neutral intergalactic gas to an IGM which is (almost) fully ionised in
its hydrogen component by $z \sim$ 6 (e.g. \citealt{Fan2006}), while helium
reionisation is believed to be complete at  $z \sim$ 2.7 (e.g.
\citealt{Madau1994, Compostella2013}).

A large number of simulations have been done to understand the formation of the
first structures and the subsequent galaxy formation and evolution. Examples of
collisionless dark-matter simulations include \cite{Shaw2006},
\cite{Boylan2009}, \cite{Kim2011}, \cite{Angulo2012} and \cite{Harnois2013}, 
while e.g. \cite{Springel2005}, \cite{2014Natur.509..177V} and \cite{Schaye2014}
follow also the gas dynamics. To study radiative feedback, radiative transfer
(RT) simulations are usually used as post-processing tools for hydro codes or
N-body codes, and they form a major aspect of the study of structure formation
in general, and the IGM reionisation in particular (e.g.  \citealt{Ciardi2003,
Iliev2006, Iliev2007, Zahn2007, Trac2008, Ahn2012, Iliev2014, Graziani2015}).
Simulations where the gas dynamics and RT effects are self-consistently
accounted for have also been performed (e.g. \citealt{Gnedin1997,
Semelin2007,Gnedin2014a,Gnedin2014b,Pawlik2015}). Large scale hydro and RT
simulations require a large amount of computational and memory resources to
handle spatial resolutions spanning several orders of magnitude. In the case of
grid-based codes, this problem has been addressed to a great extent by the use
of Adaptive Mesh Refinement (AMR) schemes (e.g. \citealt{Kravtsov1997,
Teyssier2002A&A, Miniati2007JCoPh, Bryan2014}), which provide a great amount of
theoretical and algorithmic resources, and have been used extensively to solve a
wide range of problems, such as numerical relativity, global weather and nuclear
fusion modelling (e.g. \citealt{Plewa2003}).

AMR schemes use a mesh or a grid to describe the physical domain and to
progressively increase the grid resolution in confined regions of the mesh,
based on a set of refinement criteria. By selectively increasing the resolution
only in the interesting part of the domain, AMR methods optimise the global
memory and computational resource requirements. The original idea of AMR was to
introduce a finer grid in regions of higher numerical error, which can be
identified by density gradients or by Richardson extrapolation
\citep{Berger1984484, Berger1989}. However, AMR can be employed very flexibly,
and different refinement criteria can be chosen. For example, in cosmological
applications it is customary to adopt a Lagrangian criterion in which cells are
refined when their mass is above a given threshold (e.g. \citealt{Wise2011}).
Alternatively, it is also possible to refine specific volumes in a Eulerian
style, which is advantageous for example if one is interested in resolving
turbulent motions inside cosmological halos (e.g. \citealt{Miniati2014}).

AMR is also very flexible from an algorithmic point of view. For example, it
naturally supports divergence-free magnetohydrodynamics (e.g.
\citealt{Miniati2011, Teyssier2006, Lee2009}), it has been used to make detailed
studies of thermonuclear flashes in \texttt{FLASH} \citep{Fryxell2000FLASH}, and
also to simulate the formation of large-scale cosmological structures in
\texttt{ENZO} (see e.g. \citealt{Bryan2014} and \citealt{Plewa2003} for more
examples).

Many hydro codes make use of AMR schemes to carry out gas dynamic simulations
over a range of spatial scales. Some of them are also coupled with RT schemes to
perform self-consistent simulations where the RT feedback is accounted for in
the dynamical evolution of the gas. The Hydrodynamics Adaptive Refinement Tree
(\texttt{HART}) code \citep{Gnedin2009} uses the OTVET approximation for the 3D
RT implementation \citep{Gnedin2001}, while the cosmological hydrodynamics code
\texttt{RAMSES} \citep{Rosdahl2013}, designed for simulations of structure
formation, incorporates RT using the M1 closure formalism \citep{Levermore1984}.
Both \texttt{HART} and \texttt{RAMSES} make use of the the Fully Threaded Tree
described in \cite{Khokhlov1998} to implement AMR. For the RT \texttt{ENZO} uses
an adaptive ray-tracing scheme implemented in the \texttt{HEALPix} library
\citep{AbelWandelt02, Gorskietal2005}. The interested reader can find more
examples in \cite{Iliev2009}.

Enabling AMR in a stand-alone RT code makes it suitable for post-processing the
output of many grid-based hydro codes that use the same AMR logic. By
representing the regions of interest with high resolution grids, we can follow
the details of the growth of the ionised bubbles around the sources and
understand the impact of the RT feedback effects on the surrounding environment.
A number of stand-alone RT codes implementing AMR exist: some examples are
\texttt{RADAMESH} \citep{Cantalupo&Porciani2011}, which is a Monte Carlo (MC) RT
code with a ray-tracing scheme, and \texttt{FTTE} \citep{Razoumov2005}, which
implements a scheme to perform RT on refined grids in the presence of diffuse
and point sources. A final example is \texttt{IFT} \citep{Alvarez2006}, which
has been developed to explicitly follow the I-front around a point source.

From the above examples, it is clear that AMR-enabled stand-alone RT codes are
increasingly in use and they stand to benefit from the advantages that AMR
provides. Keeping this in mind, we have developed \texttt{CRASH-AMR}, a novel
implementation of the RT code \texttt{CRASH} \citep{Ciardietal2001,
Masellietal2003, Maselli&Ferrara2005, Masellietal2009MNRAS, Pierleoni2009,
Partl2011, Graziani2013}, which is interfaced with the open source AMR library
\texttt{CHOMBO} \citep{ChomboDesign} to perform RT simulations on AMR grids. The
aim of this paper is to introduce this new version, to discuss its
implementation details and to show the tests that we have done to verify and
validate \texttt{CRASH-AMR}.

The paper is structured as follows. In Section~\ref{section:AMR-Basic} we
briefly introduce AMR and its different schemes, the \texttt{CHOMBO} library is
also presented in this context and we mention some of its applications. In
Section~\ref{section:CRASH-RT} we introduce the ray tracing implementation in
the \texttt{CRASH} code. In Section~\ref{section:CRASH-CHOMBO} we discuss the
method used to couple \texttt{CRASH} and \texttt{CHOMBO} to enable AMR in
\texttt{CRASH}. The tests performed to verify our code are discussed in
Section~\ref{section:Tests} and the results are summarised in
Section~\ref{section:Conclusion}. The advantages of the code in terms of
computational costs is highlighted in Appendix \ref{section:Grid-resolution} and
\ref{run-time}. Here, we discuss the tests done to show the dependence on
grid-resolution and performance of the code. We discuss code performance in
terms of run-time and correctness of results when compared to running RT
simulations on uniform grids. 

\section{Basics of AMR and the CHOMBO library}
\label{section:AMR-Basic}

AMR is a successful technique when a problem presents a highly inhomogeneous spatial distribution implying that 
some regions require additional resolution, i.e. additional refinement. An AMR mesh is defined as structured (SAMR) 
if its cells (typically cartesian) are connected in a regular geometry. Un-structured grids composed of triangular
or tetrahedral cells without a regular connectivity can also be defined (see for
example \citealt{Mavriplis1997}, \citealt{Khokhlov1998}, \citealt{Springel2011}
and \citealt{Paardekooper2008}). We limit our discussion to SAMR schemes in this paper.

There are many alternative ways to refine a certain subdomain using different SAMR schemes.
In the \textit{Cell-based} (CBAMR) scheme each cell of the grid is refined as and when required and 
generally a quad-tree (2D) or and oct-tree (3D) forms the hierarchical structure relating the ``coarser'' (parent) and 
``finer'' (child) cells \citep{Young19911}. The \textit{Block-structured} AMR (BSAMR) scheme tags and refines a region of the
grid, by some integer factor, based on pre-assigned criteria \citep{Berger1984484}. The various refinements are typically organised in a 
hierarchical structure connecting coarser layers (parent level) with refined layers (child level); 
the grids are then stored and maintained independent from each other \citep{Berger1984484}. When the cells that need 
to be refined are clustered together to form disjoint rectangular patches, BSAMR schemes are called
\textit{Patch-based} (PBAMR; \citealt{Dai2010}). Both BSAMR and PBAMR involve
refining a specific region in the grid rather than a single cell, and so tend to
be used synonymously in AMR-related literature.

Hereafter, we will limit our discussion to the PBAMR scheme and look at the details involved for its usage in our 
RT code \texttt{CRASH}.

To enable \texttt{CRASH} to process AMR-based grids we have adopted the
open-source AMR library \texttt{CHOMBO} \citep{ChomboDesign}. \texttt{CHOMBO}
\footnote{Chombo is a Swahili word meaning ``tool'' or ``container''.} is
actively developed at Lawrence Berkeley National Laboratory and implements a
PBAMR scheme on a C++ framework to solve systems of hyperbolic, parabolic and
elliptic partial differential equations. It has been successfully used by many
gas dynamic codes such as \texttt{CHARM} \citep{Miniati2007JCoPh},
\texttt{FLASH} \citep{Dubey2014} and \texttt{PLUTO} \citep{Mignone2012ApJS}, and
hence is suitable for our purposes as well. One could, in principle, also post-process the
outputs of codes adopting similar BSAMR frame-work as done, for example, by \texttt{ENZO} \citep{Bryan2014}. 
This would, however, require an intermediate step to 
convert both the grid hierarchy and file formats in a data format suitable for \texttt{CHOMBO}. The development of this
interface will be taken up in the future; in this paper we limit to the data that has been directly produced using the 
\texttt{CHOMBO} library or the codes adopting it, for example \texttt{CHARM}.
Test~2, in Section \ref{subsection:Test2}, shows one example of \texttt{CHARM} output post-processed by \texttt{CRASH-AMR}
\footnote{The adoption of output data produced by codes implementing 
different AMR schemes (e.g. \texttt{RAMSES}, which uses a cell-based scheme 
\citep{Khokhlov1998} could require a non-negligible effort in converting both
the grid representations and the file formats. The ray-tracing 
scheme, described in Section 4, should also sensitively adapt to the different 
nesting geometry of the AMR levels. The adoption of different AMR schemes is then 
beyond the scope of the following paper.}.

The library is organised into a hierarchy of classes, each of which provides a
specific functionality for incorporating AMR into a stand-alone code with
minimum effort, so that the software developers only need to focus on
implementing the physics. Some necessary but routine tasks, associated for
example with grid generation, management, refinement and time-stepping, are
automatically managed by the library. Hereafter, we describe some of the
functionalities implemented by \texttt{CHOMBO} that have been extensively used
in \texttt{CRASH-AMR}. For each of them, the adopted C++ class is also
indicated.

\begin{itemize}

\item \textbf{PBAMR organisation in a hierarchy of levels}. 

The PBAMR-AMR scheme implemented by the library consists of various refinement levels $L$ organized in a hierarchy starting from the base level ($L=0$), and extending up to a final level with index $L=N-1$, where $N$ is the total number of levels in the hierarchy. Each level has its own resolution defined by $r_{\mathrm{0}} \cdot r^{\mathrm{N}}$, where $r_{\mathrm{0}}$ is the resolution of level $L=0$ and $r$ is the ``refinement ratio'', i.e. the ratio of the resolution between two contiguous levels. While the entire AMR scheme is represented by a C++ class \textit{AMR}, each refinement level $L$ in the AMR hierarchy is implemented in \texttt{CHOMBO} by the class \textit{AMRLevel}. Each \textit{AMRLevel} class implements pointers to its relevant parent in the hierarchy (i.e. level $L-1$) and its child (i.e. $L+1$) to allow an easy traversal of the entire hierarchy of refinement levels.

\item \textbf{Representation of each level as composition of cell boxes}.

In the PBAMR scheme of \texttt{CHOMBO} each level is composed by cells which should be seen as minimum units of space with assigned width. Cells are organized in rectangular grids called ``boxes'' each of which occupies a unique location in the level and it is disjoint from the others, i.e. a cell at a particular refinement level can belong to only one box. \texttt{CHOMBO} represents these boxes by the \textit{Box} class. Each box occupies a unique location in the 3D space, and its 
coordinates are provided by the \textit{smallEnd} and \textit{bigEnd} functions of the \textit{Box} class. A refinement level is represented as an array of disjoint boxes (\textit{DisjointBoxLayout}) which can be traversed with iterator classes (e.g., the \textit{DataIterator} class) to access a single box. 

Boxes that lie adjacent to a box at the same refinement level are said to be its neighbors and can be accessed through the \textit{NeighborIterator} class. Subset of boxes can also be grouped together in an array called ``Fortran Array Box'' (corresponding to the class \textit{FArrayBox}) to allow an easy and fast access to the subset data for retrieval or update operations.

\item \textbf{Interactions between different refinement levels}. 

When a coarsening or refinement is requested at some level $L$, it operates on certain box(es) with a given refinement ratio $r$, also implying that the box(es) can have multiple child boxes (i.e. at level $L+1$) and can lie over multiple parent boxes (i.e. at level $L-1$). The boxes at level $L-1$ or $L+1$ whose intersection 
with a box at level $L$ is non-empty are its parent and child boxes, respectively.

When physical quantities representing continuous fields are refined in space (i.e. the result is stored across different levels) the continuity of their gradients must be ensured. For this reason the AMR scheme of \texttt{CHOMBO} adopts interpolation and averaging methods at the interface of grids. A typical example of these smoothing operations occur when the data on the grids are loaded to set up the initial
conditions (ICs) of a hydro simulation. To initialise the finer grids from the existing coarse grids, the \textit{FineInterp} class is used, while to update the coarse grids with the data on the finer grids the \textit{CoarseAverage} class is adopted.

\item \textbf{Data storage and grid I/O}. \texttt{CRASH-AMR} adopts the HDF5
data format standard\footnote{\url{http://www.hdfgroup.org/}} to store the RT
results, so that they can be easily post-processed and visualised by using
state-of-the-art visualisation software like
Visit\footnote{\url{https://wci.llnl.gov/codes/visit/home.html}} or
Paraview\footnote{\url{http://www.paraview.org/}}.

\end{itemize}

\section{Radiative Transfer code CRASH}
\label{section:CRASH-RT}

\texttt{CRASH} is a 3D MC RT code that can self-consistently follow the
formation and evolution of ionised regions created by sources present in a
static and inhomogeneous gas environment; the gas consists of H, He and metals.
The temperature evolution of the gas is calculated self-consistently.
Additionally, the code can account for an arbitrary number of point sources, as
well as a UV background. 

Our work is based on \texttt{CRASH} version 3 (\texttt{CRASH3}; refer to
\citealt{Graziani2013} and references therein for more details), and the
developments presented here contribute to \texttt{CRASH-AMR} by using
\texttt{CRASH3} as baseline but, for simplicity, without the inclusion of
metals. In this section, we discuss the \texttt{CRASH} code briefly, and we
provide only the details relevant to the AMR implementation. 

\texttt{CRASH} works by assigning the ICs onto a static, regular 3D grid which
specifies the gas number density $n_{\mathrm{gas}}$, temperature $T$, the H, He
ionisation fractions ($x_{\mathrm{HII}}, x_{\mathrm{HeII}},
x_{\mathrm{HeIII}}$), the source coordinates, luminosity $L$ and spectral energy
distribution (SED) $S$. The radiation from each source is discretised into
photon-packets represented by $N_{\nu}$ frequency bins, each containing
$N_{\mathrm{p},\nu}$ photons as determined by the SED, which are propagated
along the rays casted in random directions from the point sources. The
simulation proceeds by emitting photon-packets from all the sources and
propagating them along the rays until the end of the simulation time. 

We discuss briefly the ray tracing routine of \texttt{CRASH} in the following
paragraphs. Consider a ray along which a packet propagates by crossing a series
of cells. For each cell ${l}$ that is crossed, \texttt{CRASH} calculates the
casted path $\delta_{\mathrm{l}}$ and the corresponding optical depth of the 
cell as

\begin{eqnarray} \label{eq:castpath} 
\begin{aligned}
\tau &= \tau_{\mathrm{HI}}+ \tau_{\mathrm{HeI}}+ \tau_{\mathrm{HeII}} \\
&=[\sigma_{\mathrm{HI}}(\nu) n_{\mathrm{HI}} + \sigma_{\mathrm{HeI}}(\nu)
n_{\mathrm{HeI}} + \sigma_{\mathrm{HeII}}(\nu) n_{\mathrm{HeII}}] \delta_{l},
\end{aligned}
\end{eqnarray}
where $n_{A}$ and $\sigma_{A}$ are the number density and cross section of the
absorber $A= $ H\(\ions{I} \), He\(\ions{I} \), He\(\ions{II} \). If the packet
reaches the cell with a photon content of ${N_\gamma}$, then the number of
photons absorbed in the cell is given by

\begin{equation} \label{eq:absphot} 
 {N_\gamma^l} = {N_\gamma} ( 1 - e^{-\tau}).
\end{equation}
${N_\gamma^l}$ is then used to calculate the ionisation, recombination fractions
and temperature equations that regulate the physical state of the gas (Section~2.4
of \citealt{Masellietal2003}). The angular direction of the ray and coordinates
of the current cell are used to calculate the coordinates of the next cell that
the ray will cross; this is repeated until the photon content in the packet is
extinguished or, if periodic boundary conditions are not applied, the packet
exits the grid. We finally note that since we use \texttt{CRASH-AMR} in a
post-processing mode, operations of data smoothing, described in the section 
above, are confined to the initialisation of the RT and do not impact the ray 
tracing algorithm.

\section{Coupling CRASH and CHOMBO}
\label{section:CRASH-CHOMBO}

In this section we describe how \texttt{CRASH} and \texttt{CHOMBO} are coupled
to implement \texttt{CRASH-AMR}; we discuss both the adopted methodology and the
solutions we found to the various technical issues that occurred during the code
development.

\texttt{CRASH-AMR} is implemented in Fortran 2003, while \texttt{CHOMBO} is a
C++ code. We have created a C interface between Fortran and C++ to allow
\texttt{CHOMBO} to communicate and share information with \texttt{CRASH-AMR} by
using the interoperability features implemented in the programming languages
specifications. As discussed in Section~\ref{section:AMR-Basic},
\texttt{CRASH-AMR} uses the grid representation of \texttt{CHOMBO} (i.e. the
\textit{Box} class and its subclasses) to store the physical variables that have
a spatial representation, e.g. $n_{\mathrm{gas}}, x_{\mathrm{HII}},
x_{\mathrm{HeII}}, x_{\mathrm{HeIII}}$ and $T$. 

In the ray tracing algorithm implemented in \texttt{CRASH-AMR} (see Sec.~\ref{section:CRASH-RT} for more details), the interaction
of radiation with matter is computed in each crossed cell by solving the
ionisation and temperature equations. This implies that when a photon packet propagates through the domain it 
can cross many refined regions containing multiple PBAMR \texttt{CHOMBO} patches. Moving through different 
AMR layers implies that the instances of the \textit{Box} classes representing collection of cells at each refinement level are continuously accessed during the travel of each photon packet 
(see Sec.~\ref{section:CRASH-RT}) to update and store the
physical quantities.

The computational cost of a continuous and inefficient access to the
\texttt{CHOMBO} library could impact the global RT performance. In fact,
depending on the chosen resolution in space and the maximum refinement level
provided by the gas dynamics simulation, a single ray could traverse a large
number of cells spanning different refinement levels. This makes the box
iteration computationally inefficient when repeated for the large number of rays
required by the MC convergence (typically larger than $10^7$). Note that this is
not the way PBAMR libraries are normally used in hydro codes to access the
information: the patch-based scheme implemented in \texttt{CHOMBO} is in fact
very efficient in the management of memory and parallel computational resources,
but provides information at the grid level instead of at the cell-based
quantities, as required by the \texttt{CRASH-AMR} RT scheme. As further
complication, a realistic RT simulation generally involves an irregular
distribution in space of the sources from which a large number of rays is
emitted in random directions, implying that the boxes at each refinement level
are not accessed contiguously. Consequently, the standard interface provided by
the \texttt{CHOMBO} library cannot be simply re-used in \texttt{CRASH-AMR}. To
resolve this issue we have developed a novel Fortran data structure in
\texttt{CRASH-AMR}, minimising the run-time overhead to access and iterate the
AMR layers, as described below. 

During each ray traversal the photon-packet information has to be 
propagated through the AMR grid hierarchy. If we assume that the source lies in the highest 
refinement level, then the ray 
will propagate, starting from that level, to the coarser levels below and move 
back to finer levels if present. At each step of the RT simulation, the data 
structure needs to know the refinement level a cell belongs to, whether the cell 
is refined or not, and which box contains the refined cell; the way all these 
quantities are accessed
and the mapping between \texttt{CRASH} and  \texttt{CHOMBO} data is described in
the following paragraphs. Hereafter, we will refer to the grid with the lowest 
resolution as `base grid', while the refined grids will be referred to as 
`refined levels'.

\begin{figure}
   \centering
   \includegraphics[width=8.3cm, height=5.5cm]{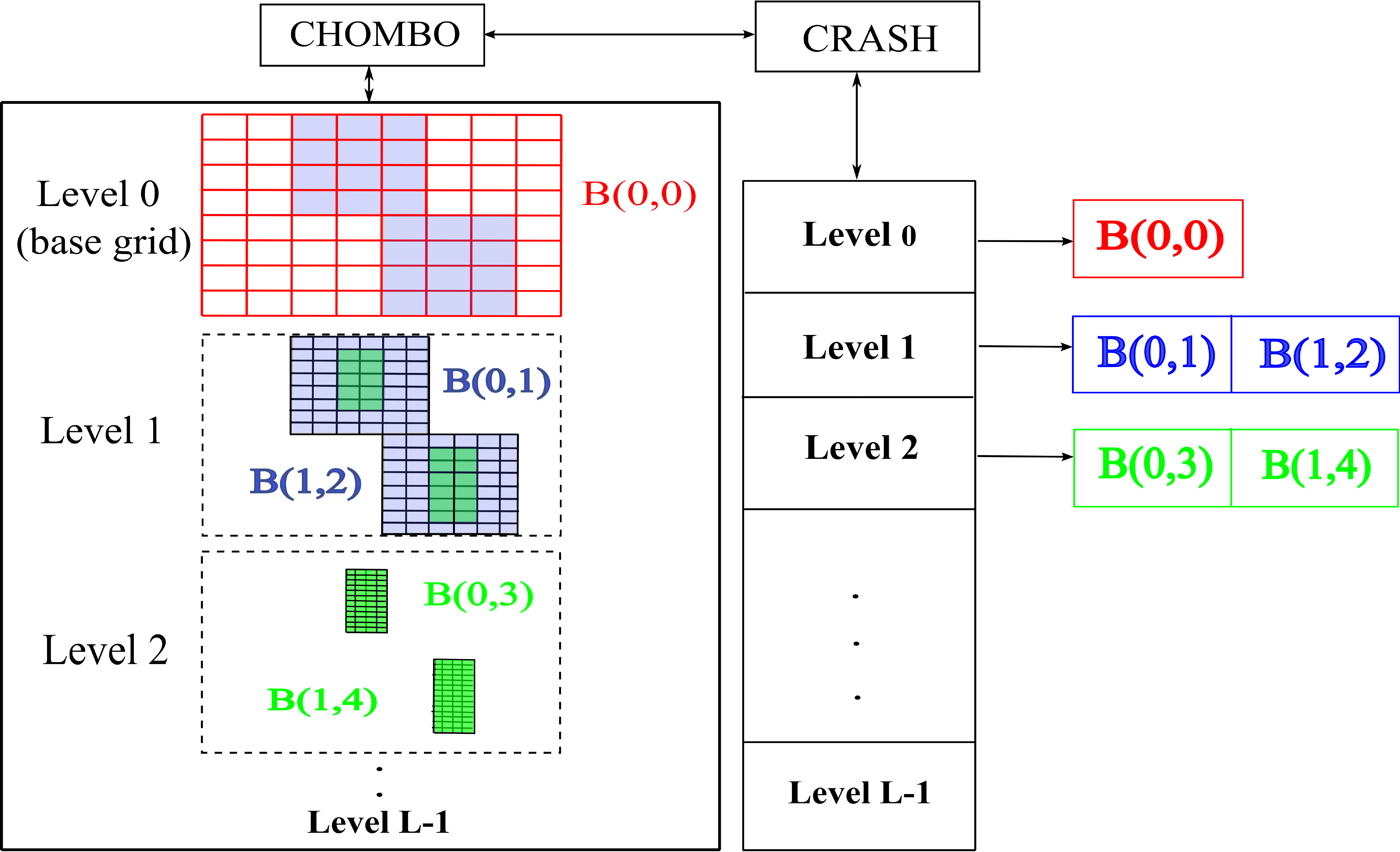}
   \caption{Interface between \texttt{CRASH} and \texttt{CHOMBO}, the grid
hierarchy in \texttt{CHOMBO} being reflected through the data structure built in
\texttt{CRASH}. The AMR grids, stored as an array of boxes in \texttt{CHOMBO},
are stored in \texttt{CRASH} on a level basis, with pointers to the grid data at
each level. Each box has an associated \textit{localID} and \textit{globalID}.
See text for more details.}
   \label{fig:CRASH-CHOMBO}
\end{figure}

Figure~\ref{fig:CRASH-CHOMBO} shows the data representation in \texttt{CHOMBO}
on the left-hand side (see also Section ~\ref{section:AMR-Basic} for more details),
and the \texttt{CRASH} equivalent on the right-hand. We also use similar colours
in both sides to represent corresponding boxes at a given refinement level. A
simplified but representative \texttt{CHOMBO} hierarchy, consisting of the base
level 0 and its refinements from 1 to $L-1$, is shown in the picture; for
clarity purposes we just represent the boxes at levels 0, 1 and 2. To help the
reader in connecting this picture with the abstract data representation provided
in Section \ref{section:CRASH-RT}, we point out that each refinement level
(dashed boxes on the left) is implemented in computer memory by an instance of
the \textit{AMRLevel} class, while the array of boxes at each level is
implemented by the \textit{DisjointBoxLayout} class. Each box, for example the
B(0,0) at base level 0 (red box on the left), is an instance of the \textit{Box}
class. 

The \texttt{CRASH} counterpart of the AMR hierarchy is mapped on the right-hand
side of Figure \ref{fig:CRASH-CHOMBO}: the base grid is represented as
refinement level 0, while the $L$ AMR grid levels are mapped with an array
containing pointers to specific properties of each box. The boxes
at each level are uniquely identified, on both sides, by an associated
\textit{localID} and \textit{globalID}. We first use the 
\textit{localID} to get direct access to the right box in the \textit{DisjointBoxLayout} array, 
and then to the physical variables of interest during the RT simulation. The 
corresponding \textit{globalID} along with the refinement level provides an 
index into one of the arrays on the \texttt{CRASH} side. For example, box B(0,1) 
at refinement Level 1 has a \textit{localID} of 0 and a \textit{globalID} of 1. 
The value of the \textit{localID} indicates that B(0,1) is at position 0 of the 
disjoint box array storing all the boxes at that level, and the physical
data can be finally accessed through its \textit{FArrayBox} (see
Sec.~\ref{section:AMR-Basic} for more details on the classes). The
start and end coordinates of the box, which determine its size, are also stored
to allow the ray tracing algorithm to recognise if a ray has exited a box at a 
given refinement level. Along with the physical data, we store in each cell 
the \textit{globalID} of the box it belongs to. For cells that are covered by 
the cells of a refined box, we store the \textit{globalID} of the refined box. 
This is used to determine if a cell that the ray is passing through is refined 
or not. Additionally, since a PBAMR scheme allows a refined box to lie over 
multiple coarse boxes, we keep a list of all parents. This is done by storing, 
for each box, the \textit{globalIDs} of its parent(s). Finally, we also store a 
neighbor list containing the \textit{globalIDs} of all boxes that are 
neighbor(s) to a box.

\begin{figure*}
   \centering
   \includegraphics[width=15.3cm, height=13.0cm]{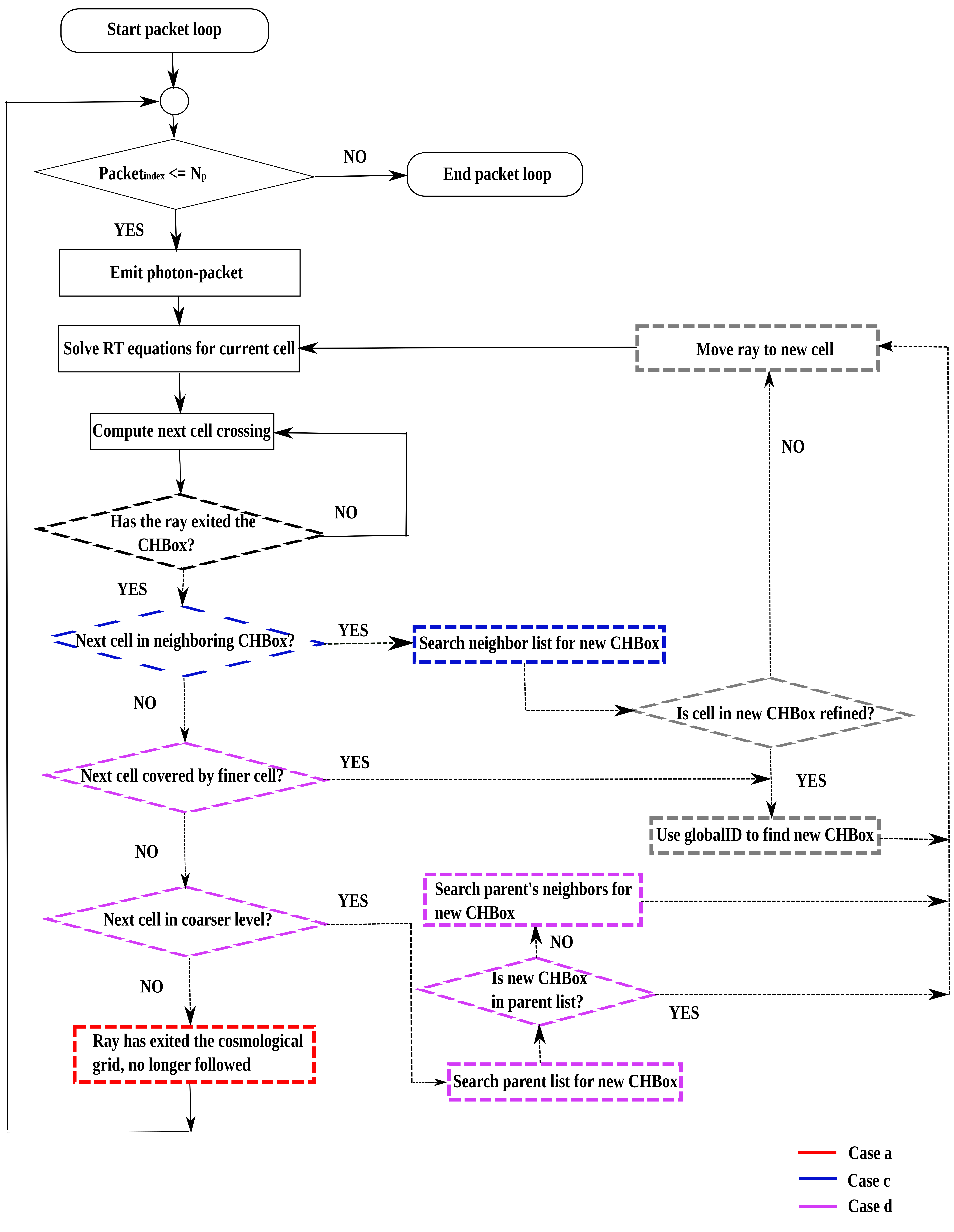}
   \caption{\texttt{CRASH-AMR} flowchart describing the interplay between ray tracing and 
            the \texttt{CRASH}-to-\texttt{CHOMBO} interface when a photon packet travels 
            through many AMR levels. The boxes with solid lines indicate the \texttt{CRASH}-side of the algorithm,
            while those with dashed lines indicate the \texttt{CHOMBO} interface-side. The solid/dashed lines connecting 
            the boxes indicate if the \texttt{CRASH}/\texttt{CHOMBO} interface-side of the algorithm, respectively, is being called.
            Various colours refer to the different algorithmic scenarios that could happen during a photon 
            packet propagation. See text for more details.}
   \label{fig:CRASH-CHOMBO-WORKFLOW}
\end{figure*}

The following paragraphs describe how the new data structure is used during the
ray tracing algorithm. First note that the sources emitting photons do 
not move
across the grid during a RT simulation. Hence, given the refinement level the 
source lies in and its coordinates we need to look for the \textit{globalID} of 
the box, containing the source, only once. This can then be used to index into 
the arrays in \texttt{CRASH}, to get the right box and the physical data stored 
at the source coordinates. Then, during the propagation of the photon packet, at 
each cell crossing, the following scenarios apply:

\begin{enumerate}[(a)]
 \item \label{ray1} the ray might escape the grid and then we no longer follow
it unless periodic boundary conditions are applied;
 \item \label{ray2} the photon content of the packet is completely absorbed and
then the propagation stops;
 \item \label{ray3} the ray crosses the cell and enters a new cell at the same
AMR level;
 \item \label{ray4} the ray crosses the cell and enters another cell at a finer
(or coarser) AMR level.
\end{enumerate}
While cases (\ref{ray1}) and (\ref{ray2}) do not need further comments, for case
(\ref{ray3}) the new cell might lie in the same box or it might enter a new one.
In the former case, we just continue the ray propagation as described 
earlier;
in the latter case, we use the coordinates of the new cell the ray is in and 
search the  
neighbors of the box that the ray was previously in, at the same refinement 
level, for the box that contains the new cell. Finally,
case (\ref{ray4}) needs a different approach because the cell optical depth,
calculated using the casted path, depends on the refinement level the ray is
crossing through (see Eq. \ref{eq:castpath}). Here, again, different scenarios
apply:

\begin{enumerate}[(1)]
 \item \label{child} the ray enters a finer level. Given the new cell 
coordinates, we check if the \textit{globalID} stored in the cell is the same as 
the \textit{globalID} of the box the ray was in. If not, this indicates that the 
cell is covered by a refined cell. The \textit{globalID} stored in 
the new cell is then used to find the refined box and the corresponding cell. 
Finally the ray is moved to the finer level;
  \item \label{parent} the ray enters a coarser level. We search the 
parent list, of the box the ray was previously in, for the new box containing 
the new cell.
 \end{enumerate}

Once the new box is found in the neighbor list for case (\ref{ray3}) or 
in the parent list for (\ref{parent}) we
recursively move the ray, as in case (\ref{child}), to a finer level if the 
neighbor or parent is refined.   
The same procedure is repeated until cases (\ref{ray1}) and/or (\ref{ray2})
apply, and the photon packet propagation stops.  

Figure~\ref{fig:CRASH-CHOMBO-WORKFLOW} shows the logical workflow described above.
 In this diagram the solid lines of the flowchart indicate the standard 
 steps of the \texttt{CRASH} ray tracing algorithm, while the dashed lines indicate the 
 steps performed by the \texttt{CRASH}-to-\texttt{CHOMBO} interface to move across the various
  \texttt{CHOMBO} boxes (CHBox) of the AMR side. The many scenarios described above 
  can be visually followed by colours, as shown in the legend of the figure.
 
It is important to note that, although we use \texttt{CHOMBO} to initialise and
store the AMR grid, once this data has been mapped onto the \texttt{CRASH} side,
our implementation does not call any \texttt{CHOMBO} routines during the RT
simulation. The data structure is used purely to cross the levels and have a
fast access to the grid data. As a consequence of this architectural choice,
there is no overhead of using \texttt{CHOMBO} during the ray-tracing routine but
the time needed to find the right cell. 

As a final consideration, we want to emphasise that the above features are
included as a separate functionality, allowing the user to enable or disable the
use of AMR grids to do RT simulations and use the traditional data storage
setting up the simulation ICs only on the base grid. If the AMR functionality is
enabled, then the user can either run simple tests with pre-defined refinement
criteria, or more realistic cases with AMR grids provided by hydro codes, as
shown in the following section. By adopting pre-defined refinement criteria, the
user could decide, for example, to refine an arbitrary part of the base grid and
set up specific test cases, while for realistic gas configurations, the
refinement is generally determined by the hydro code, and \texttt{CRASH-AMR}
operates in post-processing mode.

\section{Tests and Results}
\label{section:Tests}

In this section we show the results of some of the tests we have performed to
guarantee the reliability of the new AMR implementation. We run a number of test
cases in idealised configurations; in Test~1 we compare \texttt{CRASH-AMR} with
the AMR functionality disabled to \texttt{CRASH3}, and then we compare results
with/without the AMR functionality enabled. These set-ups are useful to check
the numerical noise introduced by the presence of the AMR grids on the RT
algorithm. In Test~2, we apply \texttt{CRASH-AMR} to a realistic density field
from the \texttt{CHARM} simulations described in \cite{Miniati2007JCoPh}.
Additionally, in Appendix \ref{run-time} we take a further look at the
performance of the code in terms of run times and correctness of results. 
Henceforth, we use $d$ to represent the comoving distance from a point source,
the units in kpc or Mpc are indicated accordingly.

\begin{figure} 
  \subfloat{\includegraphics[width=\columnwidth, height=14cm]{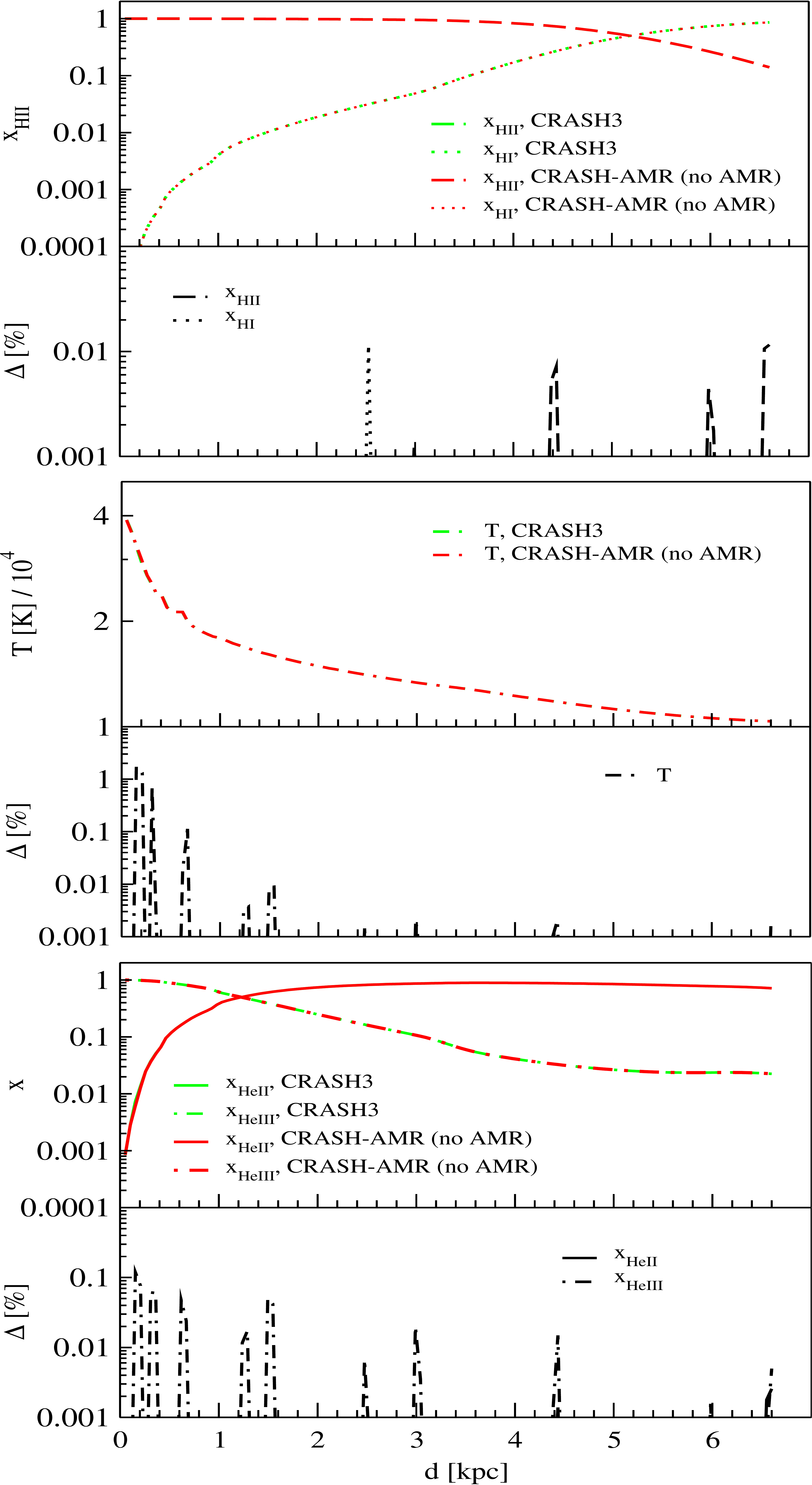}}
  \setlength{\abovecaptionskip}{10pt}
  \caption{Spherically-averaged profiles at time $t=500$~Myr for Test~1a. The
colours refer to \texttt{CRASH3} (green) and \texttt{CRASH-AMR} (AMR disabled;
red). The bottom sub-panels show $\Delta$ between the \texttt{CRASH3} and
\texttt{CRASH-AMR} (AMR disabled) results. \textbf{Top:} Profiles of
$x_{\mathrm{HII}}$ (dashed lines) and $x_{\mathrm{HI}}$ (dotted lines).
\textbf{Middle:} Profile of $T$ (dash-dot lines). \textbf{Bottom:} Profiles
of $x_{\mathrm{HeII}}$ (solid lines) and $x_{\mathrm{HeIII}}$ (dot-dash
lines).} 
  \label{fig:spherical-T1}
\end{figure}

\subsection{Test~1: Str\"omgren sphere in a H+He medium}
\label{subsection:Test1-AMR}

We have set up a test equivalent to Test~2 of the Radiative Transfer Code
Comparison Project (RTCCP; \citealt{Iliev2006a}). The test simulates the
evolution of an ionised region around a single point source located at the grid
origin (1,1,1) in a box of side length $L=6.6$~kpc and
mapped on a grid of $128^{3}$ cells. The source is assumed to be steady with an
ionising rate of $\dot{N_{\mathrm{\gamma}}}$ = $5 \cdot 10^{48}
\mathrm{photons}\cdot \mathrm{s^{-1}}$ and an associated black-body spectrum at
temperature $T_{BB} = 10^{5}$~K. The volume is filled by a uniform and static
gas of number density $n_\mathrm{gas}$ = $10^{-3} cm\mathrm{-3}$,
containing H (92\% by number) and He (8\%). The gas is assumed to be fully
neutral and at a temperature $T$=100~K, which is then calculated
self-consistently with the progress of ionisation for a simulation time
$t_\mathrm{sim}=500$~Myr, starting at redshift $z$ = 0.1. We output the results 
at intermediate times $t$ = 10, 50, 100 and 200~Myr as in the original set-up. It
should be noted that simpler tests (e.g., with a gas composed by H only or with
the temperature kept constant) have been run as well, and give results similar
to those discussed in the following. 

\subsubsection{Test 1a: AMR disabled}
\label{subsubsection:Test1-noAMR}

To verify that the changes done to enable RT on AMR grids do not introduce any
numerical noise, we have run Test~1 with AMR disabled in \texttt{CRASH-AMR} and
compared the results to those from \texttt{CRASH3}.

The outcome is given in Figure~\ref{fig:spherical-T1}, where each panel shows
spherically-averaged physical quantities as a function of $d$, together with the
percentage difference ($\Delta$) between the \texttt{CRASH3} and the
\texttt{CRASH-AMR} results. We define $\Delta = (R_\mathrm{ref} - R_\mathrm{i}) 
\cdot 100 ~/ \
R_\mathrm{ref}$, where $R_\mathrm{ref}$ and $R_\mathrm{i}$ refer to results of 
\texttt{CRASH3} and
\texttt{CRASH-AMR}, respectively. From the Figure it is clear that there is no
significant difference between the two codes, while the spikes that we see in
$\Delta$ (with a maximum of $\sim$1\%, but mostly below 0.1\%) are due to
numerical artefacts caused by optimisation of \texttt{CRASH-AMR} involving
rearranging of double-precision floating point arithmetic expressions, 
and are not due to the
changes associated with \texttt{CHOMBO}. This shows that the AMR feature in
\texttt{CRASH-AMR} is isolated from the rest of the code and can be disabled
without introducing any numerical noise into the results. 

\begin{figure}
  \subfloat{\includegraphics[width=\columnwidth, height=14cm]{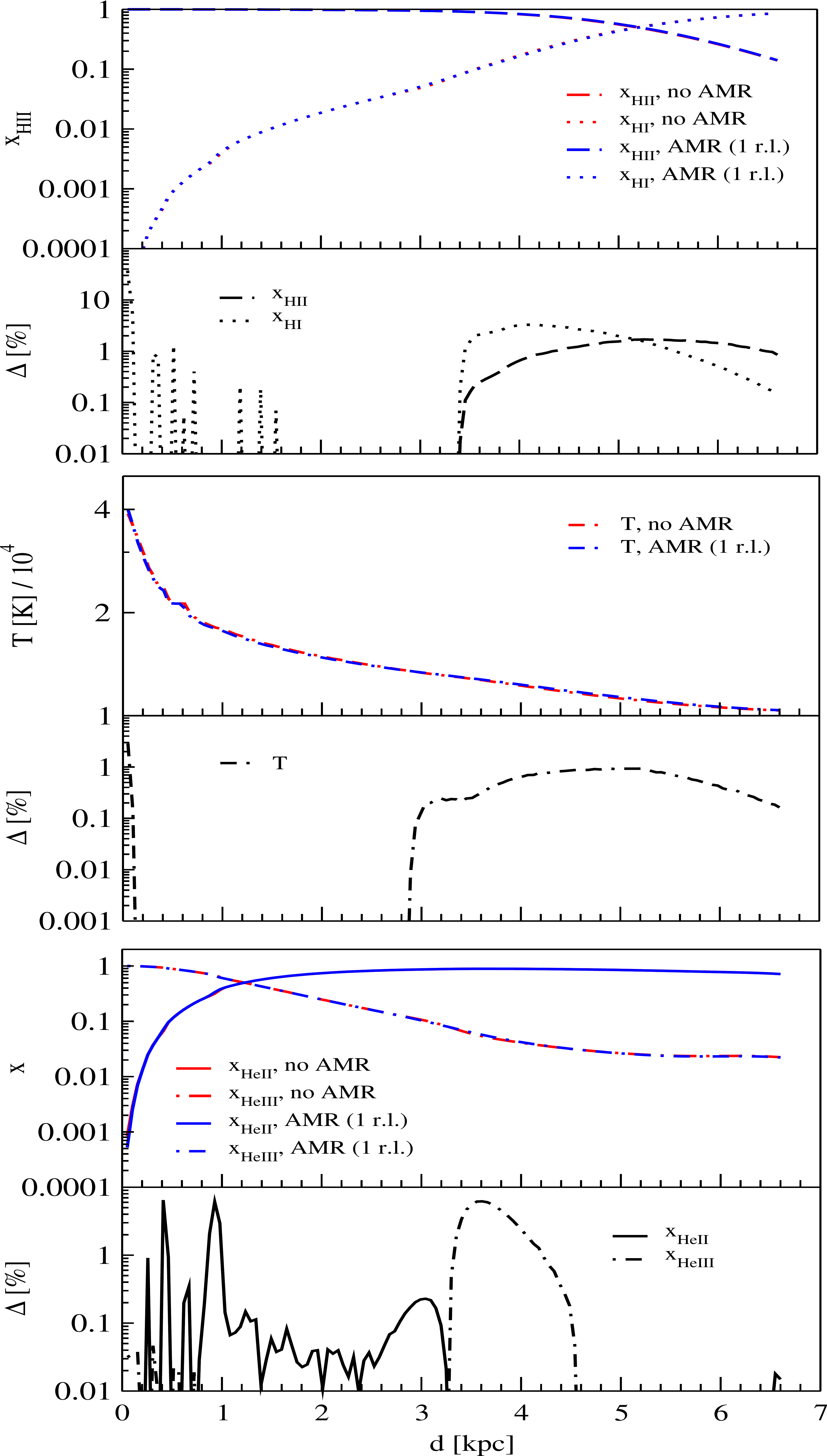}}
  \setlength{\abovecaptionskip}{10pt}
  \caption{Spherically-averaged profiles at time $t=500$~Myr for Test~1b. The
colours refer to \texttt{CRASH-AMR} (AMR disabled; red) and \texttt{CRASH-AMR}
with one refinement level (1 r.l.; blue). The bottom sub-panels show $\Delta$
between the \texttt{CRASH-AMR} (AMR disabled) and \texttt{CRASH-AMR} (1 r.l.)
results. \textbf{Top:} Profiles of $x_{\mathrm{HII}}$ (dashed lines) and
$x_{\mathrm{HI}}$ (dashed lines). \textbf{Middle:} Profile of $T$ (dash-dot
lines). \textbf{Bottom:} Profiles of $x_{\mathrm{HeII}}$ (solid lines) and
$x_{\mathrm{HeIII}}$ (dot-dash lines).} 
  \label{fig:spherical-T1-AMR}
 \end{figure}

\subsubsection{Test 1b: AMR enabled}
\label{subsubsection:Test2-AMR}

Our next step has been to run the previous test with AMR enabled, refining
$100^\mathrm{3}$ cells around the source by a factor of 2, with the ICs set up on the base grid as 
well
as on the refined grid. The source is now surrounded by a cartesian grid, at an equivalent resolution of 
$256^\mathrm{3}$, of $\sim5~kpc$. The results are given in Figure
\ref{fig:spherical-T1-AMR}, where each panel shows spherically-averaged physical
quantities as a function of $d$, together with the $\Delta$ between the results
from \texttt{CRASH-AMR} (AMR disabled, $R_\mathrm{ref}$) and \texttt{CRASH-AMR} 
with
one refinement level (1 r.l., $R_\mathrm{i}$). Here we find that $\Delta$ can be 
as
high as 10\% for $x_{\mathrm{HI}}$ and $x_{\mathrm{HeIII}}$ cells very close to
the source ($d\sim$ 0 - 1 kpc), and at a distance of 3 - 6 kpc, in the partially
ionised region. More typical values, though, do not exceed 1\% for all the
physical quantities.

From Section \ref{subsubsection:Test1-noAMR} we know that, with the exception of
the spikes observed for $x_{\mathrm{HI}}$, the differences seen are not due to
the implementation in \texttt{CRASH-AMR}, but they must rather be associated to
the higher grid resolution in the refined region. Also, these differences are not 
due to any spurious noise produced by this particular method of calculating the average, but is rather
due to the higher resolution being used. When we have a grid at a
higher refinement level all the refined cells that cover a coarse cell might not
lie within a given radius from the source and thus not contribute to the
spherical average. This effect is more prominent in the partially ionised regions where two adjacent cells 
might not have the same e.g. $x_{\mathrm{HII}}$ values, unlike a fully ionised region where $x_{\mathrm{HII}}$ is 1. 
If a cell does not lie within the given radius, the average value will differ. 
Also note that Helium has a recombination rate five times higher than that of H. As a result,
the helium components show more sensitivity and we get differences of $\sim$ 1 - 10\% in many cells.
\texttt{CRASH-AMR} will then provide a better description of the regions with
fully ionised helium, generally more confined to the brightest and x-rays
luminous sources (e.g. quasars). 

Additional tests showing the dependence of the results on the grid resolution
are detailed in Appendix \ref{section:Grid-resolution}. 

\subsection{Test~2: a realistic density field}
\label{subsection:Test2}

In this section we apply \texttt{CRASH-AMR} on a density field snapshot obtained
from a simulation run within the Santa Barbara Cluster Comparison
Project, where the formation of a galaxy cluster in a standard CDM universe is
followed \citep{Frenk1999}. The simulation has been performed by the hydro code
\texttt{CHARM} \citep{Miniati2007JCoPh} in a box size $L=64$~Mpc (comoving) at
redshift $z=0.1$. The cosmological parameters are $\Omega_\mathrm{m}=1,
\Omega_\mathrm{b}=0.1$, $\Omega_\mathrm{l}=0$ and 
$H_\mathrm{0}$=50~km~s$^\mathrm{-1}$~Mpc$^\mathrm{-1}$. The
simulation is initialised at $z=40$ with a base grid of $64^\mathrm{3}$ cells
representing a box of $64^\mathrm{3}$~Mpc$^\mathrm{3}$ comoving, and a grid
of $128^\mathrm{3}$ cells placed at the centre of the base grid and representing 
a box
of comoving length $32$~Mpc. Only the central region is refined based on a
local density criterion, with a refinement ratio of 2. At the end of the
simulation there are six refinement levels in total, along with the base
grid. The cell width at the coarsest level is 1~Mpc, while that at the finest
level, with an equivalent resolution of $4096^\mathrm{3}$ cells, is 15 
kpc. The code \texttt{CHARM}
adopts the \texttt{CHOMBO} library to implement the AMR functionality, and the
HDF5 files available from the output of this simulation can be immediately used
as an input to \texttt{CRASH-AMR} by extracting the necessary information from
the HDF5 metadata. As the simulation does not provide information on the star
formation, we define the point source locations associating them to the gas
density peaks at the most refined level. 

We set up the following RT simulations: 

\begin{enumerate}[(a)]
 \item \label{PS-spread} multiple point sources placed at the highest refinement
level, at locations far enough so that when moved to lower refinement levels
they do not gather. The sources are monochromatic, the gas temperature is kept
constant throughout the simulation;
  
 \item as (\ref{PS-spread}), but now the point sources are placed at locations
close enough so that they can be gathered at the lower refinement levels; 
 
 \item same point source locations as (\ref{PS-spread}), but with a black-body
spectrum, and the gas temperature is calculated self-consistently with the
progress of ionisation during the simulation.
\end{enumerate}

In all cases the point sources are located within high density peaks, chosen to
ensure that the criteria mentioned above are satisfied for our tests. 
Since the base grid is refined only in the central 
region within which the higher refinement levels also lie, the resolution
at large distances from the point sources is the same for all cases.

We set a reference ionisation rate, \textit{$\dot{N}_{\gamma,ref}$}, for the
source in the highest gas density peak. For the other sources $i$:
\begin{equation}
\label{N_ref}
 \dot{N}_{\gamma,i} = \frac{\dot{N}_{\gamma,ref} \cdot m_\mathrm{i}} 
{m_\mathrm{ref}} ,
\end{equation}
where $m_\mathrm{ref}$ and $m_\mathrm{i}$ are the mass in the cell containing 
the reference
source and source $i$, respectively. The initial temperature is $T=100$~K and
the
gas is assumed to be fully neutral. The simulation time is 500~Myr.

To emphasise the advantage of an AMR scheme, we compare results of simulations
run with different refinement levels. Additionally, as mentioned above, the
sources are located at the highest refinement level, so if one or more of them
lie within the same cell at the coarser levels, we consider them to be a single
source with luminosity given by the sum of the corresponding luminosities at the
finest level.

To ensure a good convergence of the MC code, we sample the radiation field with
a number of photon packets high enough to reach convergence for each
test case run at different refinement levels. We find that the MC scheme
converges with $10^\mathrm{8}$ photon packets per source ($0.07\%$ difference in 
volume
averaged $x_{\mathrm{HII}}$ values between two test cases with $10^\mathrm{8}$ 
and
$10^\mathrm{9}$ photon packets per source). However, the convergence of the MC 
scheme
is very much problem dependent, hence we do not discuss this further.           
 
\begin{figure*} 
 \subfloat{\includegraphics[width=\textwidth]{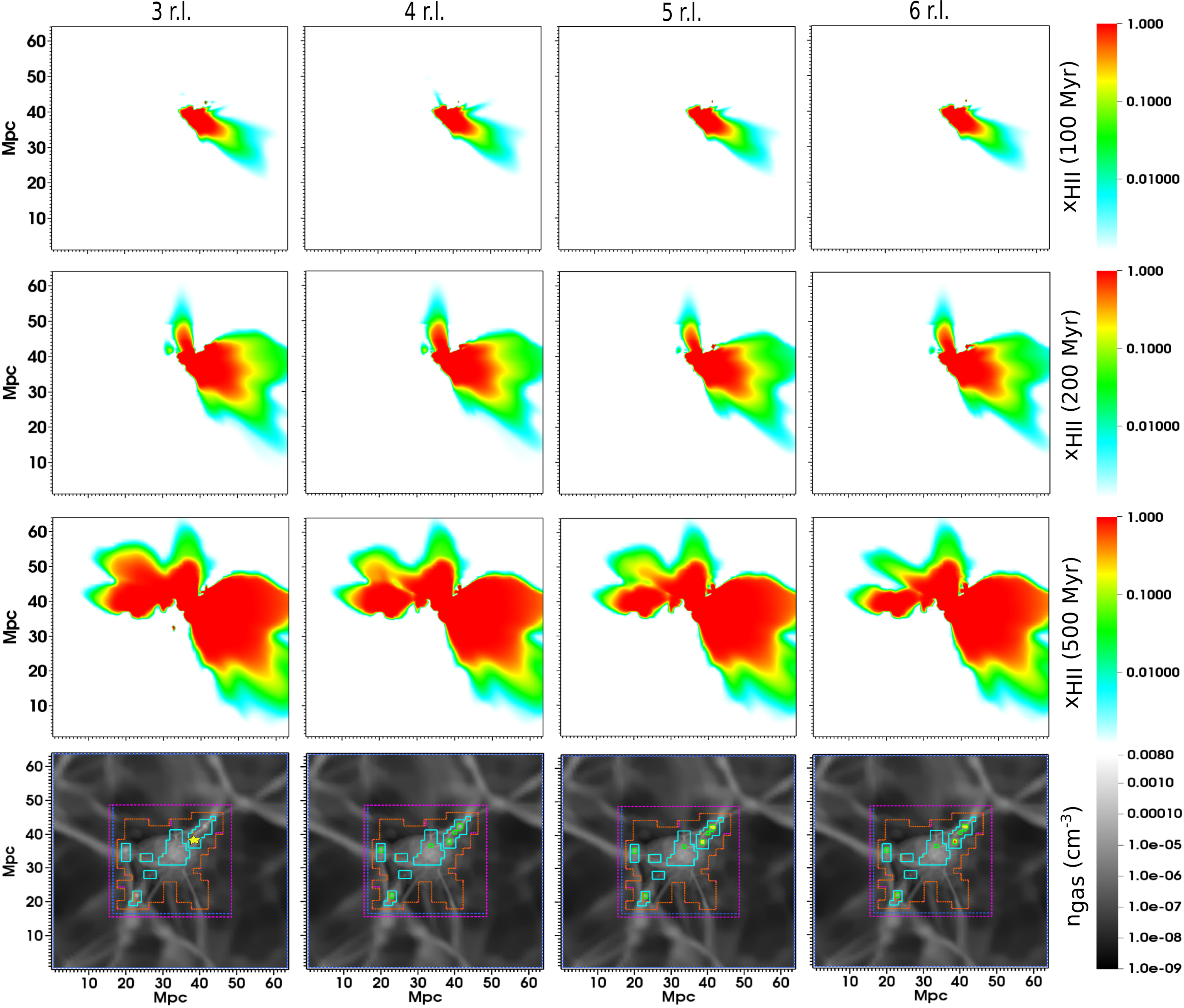}}
 \setlength{\abovecaptionskip}{10pt}
  \caption{Maps cut through the simulation volume for Test~2a. \textbf{Top:}
Maps of $x_{\mathrm{HII}}$ at time $t=100$~Myr. \textbf{Second from top:} Maps
of $x_{\mathrm{HII}}$ at time $t=200$~Myr. \textbf{Third from top:} Maps of
$x_{\mathrm{HII}}$ at time $t=500$~Myr. \textbf{Bottom:} Maps of
$n_{\mathrm{gas}}$, the dotted lines represent the extent of the different
refinement levels associated with $n_{\mathrm{gas}}$ (the base grid is not seen
here): magenta (1st r.l.), orange (2nd r.l.), cyan (3rd r.l.), green (4th r.l.),
yellow (5th r.l.) and red (6th r.l.). From left to right, the columns refer to
simulations run with three, four, five and six refinement levels (see text for
more details).}
\label{fig:vdensity-fsources}
\end{figure*}

\begin{figure*} 
 \subfloat{\includegraphics[width=\textwidth]{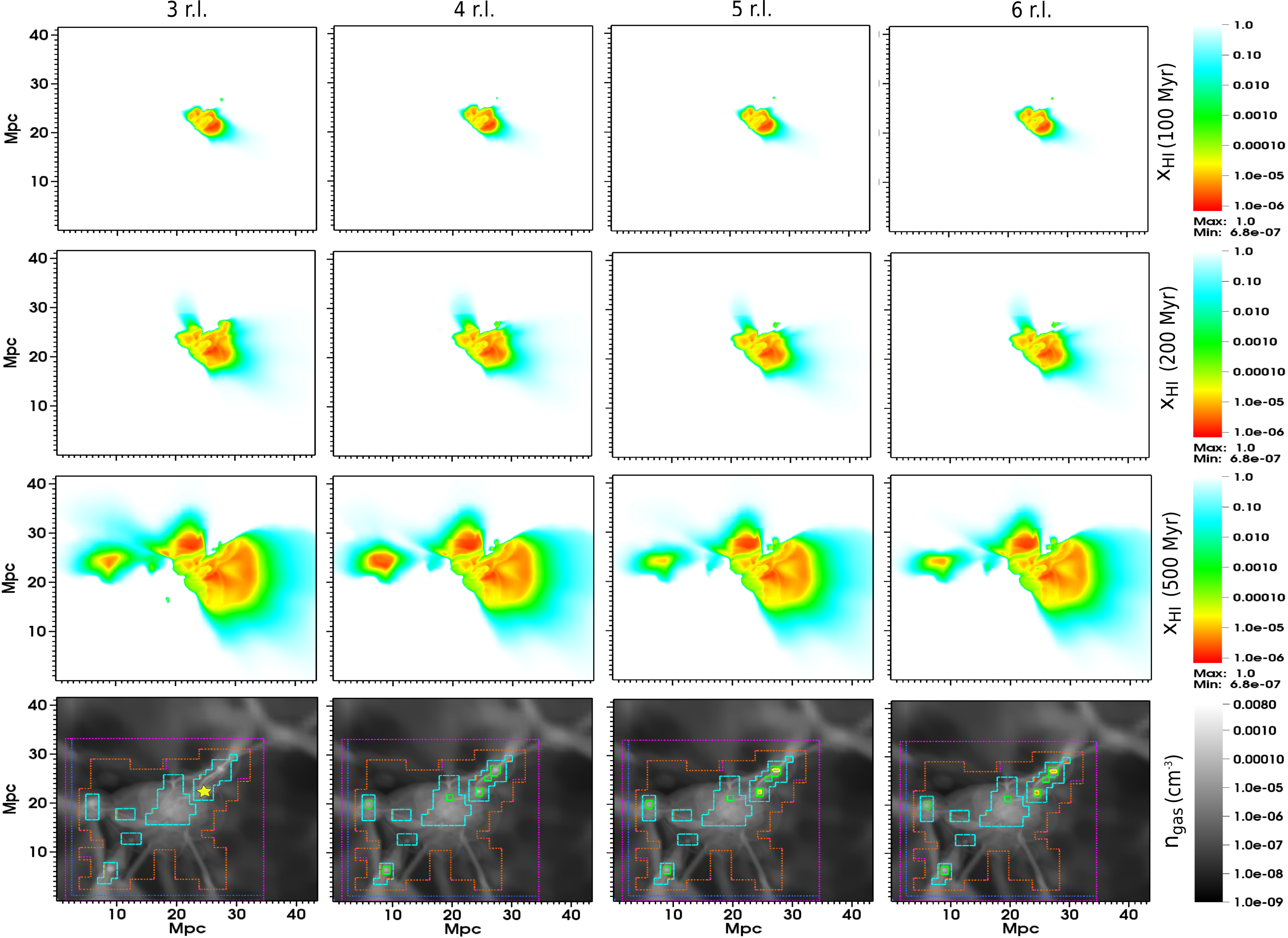}}
 \setlength{\abovecaptionskip}{10pt}
  \caption{Maps cut through the simulation volume for Test~2a. \textbf{Top:}
Maps of $x_{\mathrm{HI}}$ at time $t=100$~Myr. \textbf{Second from top:} Maps
of $x_{\mathrm{HI}}$ at time $t=200$~Myr. \textbf{Third from top:} Maps of
$x_{\mathrm{HI}}$ at time $t=500$~Myr. \textbf{Bottom:} Maps of 
$n_{\mathrm{gas}}$, the dotted lines represent the extent of the different 
refinement levels associated with $n_{\mathrm{gas}}$ (the base grid is not seen 
here): magenta (1st r.l.), orange (2nd r.l.), cyan (3rd r.l.), green (4th r.l.), 
yellow (5th r.l.) and red (6th r.l.). 
From left to right, the columns refer to
simulations run with three, four, five and six refinement levels (see text for
more details).}
\label{fig:vdensity-fsourcesXHI}
\end{figure*}

\subsubsection{Test~2a: multiple sources set far apart - constant $T$}
\label{subsubsection:vary-density-far-sources}

Here we place twenty point sources far enough from each other to remain separate
at all refinement levels. This configuration tests the effect of
grid resolution on the RT simulation. We adopt \textit{$\dot{N}_{\gamma,ref}$} =
$8 \cdot 10^{53} \mathrm{photons} \cdot \mathrm{s^{-1}}$, each point source is
monochromatic with $E_{\nu}$ = 13.6~eV, and the temperature is kept constant
throughout the simulation. For simplicity, we consider a H only gas.

Figure \ref{fig:vdensity-fsources} shows the maps of $x_{\mathrm{HII}}$ created
in simulations with different refinement levels at times $t$ = 100, 200 and
500~Myr. In the bottom panels we also show the gas number density field ($n_{\mathrm{gas}}$). 
Dotted lines represent the extent of the different
refinement levels (see the caption for more details). For reference, we also
show the location of the most luminous point source. Note that, because the RT
is done in post-processing, the gas configuration does not change during the
ionisation evolution. 

At $t=100$~Myr the $x_{\mathrm{HII}}$ maps are very similar. Differences become
more visible at $t=200$~Myr, where separate bubbles can be seen on the right
side of the box as we go to higher refinement levels. The largest differences
are present at the final time $t=500$~Myr. From a comparison between the
ionisation and gas number density maps, we can observe no direct correlation
between positions of the refinement levels and differences in the ionisation
pattern, as the sources are able to maintain their surrounding regions fully
ionised against the progressively steeper changes in density introduced by AMR. 
On the other hand, the extent of the fully and partially ionised H\(\ions{II} \)
regions shows obvious differences, as they get smaller and sharper with higher
resolution. This is due to the larger changes in density and gas recombination
rate (which increases by a factor of 3.5 between 6 and 3 r.l.) present in the
more refined grids. As a result, the escape of ionising photons becomes more
difficult, delaying the propagation of the ionisation fronts.

Finally, note that the presence of multiple point sources on different planes of
the cube and resolved by different AMR layers, creates an intricate combination
of three-dimensional RT effects in the final configuration of the overlapping
H\(\ions{II} \) fronts. This is more evident in the case with 5 and 6
refinements, in two distinct regions. The H\(\ions{II} \) region on the left is
in fact formed by a point source lying on a plane different from the one hosting
the most luminuous point source, and evolves differently with increasing number
of refinement levels. This provides a final bubble distribution and overlap in
space which is very sensitive to the number of adopted AMR refinements. 

\begin{figure*} 
 \subfloat{\includegraphics[width=\textwidth, height=4.2cm]{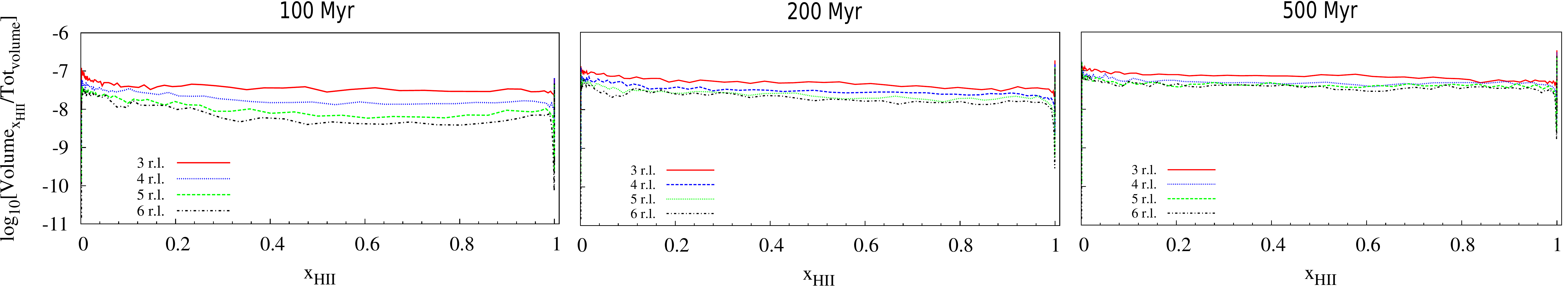}}
 \setlength{\abovecaptionskip}{10pt}
  \caption{PDFs of $x_{\mathrm{HII}}$ in the simulation volume for Test~2a. Each panel refers to a different simulation time. \textbf{Left:} Simulation time $t=100$~Myr. 
  \textbf{Center:} Simulation time $t=200$~Myr. \textbf{Right:} Final time $t=500$~Myr. The lines refer to a test case
with 3 (solid red), 4 (dotted blue), 5 (dashed green) and 6 (dot-dash black) refinement levels.}
\label{fig:vdensity-fsourcesPDFs}
\end{figure*}
 
The effects of resolving more and more gas clumps by progressively 
increasing the AMR resolution are better appreciated by showing complementary 
$x_{\mathrm{HI}} = 1-x_{\mathrm{HII}}$ maps. This is done in Figure \ref{fig:vdensity-fsourcesXHI},
  with the same panel organization of the previous figure. First note that the region 
shown here is closer to the image center; this is done to better zoom-in the spatial 
distribution of the neutral fraction at all times\footnote{As effect of the re-centering, the distance scale in Figure \ref{fig:vdensity-fsourcesXHI} does not correspond to the one in the previous figure.}. Also note that in these panels the colour palette 
indicates neutral gas fraction in logarithmic scale with a cyan-to-white 
transition corresponding to almost neutral gas, while orange-to-red areas mark 
almost ionised gas (i.e. $x_{\mathrm{HI}} \lesssim 10^{-5}$). 
As commented above, the AMR refinements close to the sources can resolve more 
over-dense structures. While the radiation is sufficient to substantially 
ionise the entire area ($x_{\mathrm{HII}} \gtrsim 0.9$) and to allow the escape of ionizing 
photons in far under-dense voids (see for example the one in the lower right side of the panels),  
many inner regions still show an intricate pattern of residual neutral gas: a large yellow area 
preserving a residual fraction $x_{\mathrm{HI}} \sim 4 \times 10^{-4}$ surrounds the red spots 
and blends into green and cyan areas when the neutral fraction progressively increases 
up to $x_{\mathrm{HI}} \sim 10^{-2}$ and $x_{\mathrm{HI}} \sim 10^{-1}$, respectively.
At megapaserc scales the structures in the various panels differ for only few, minor details,
while the external contours show a clear reduction of the ionised gas (e.g. focus on the cyan area connecting the two central H\(\ions{II} \) regions)
from left (3 r.l.) to right (6 r.l.).

Also the volume averaged ionisation fraction depends on the refinement levels
used, with $x_{\mathrm{HII}}$=4.94, 4.74, 4.51 and 4.39 $\cdot 10^{-2}$ for the
3rd, 4th, 5th and 6th refinement levels at time $t$ = 500~Myr, respectively,
with a 12.5\% difference between the test cases with 3 and 6 refinement levels.
A better statistical description of the differences induced by an increase of 
        AMR refinement levels is provided  by the probability density functions 
        (PDFs) of the ionised fraction $x_{\mathrm{HII}}$.
         In the three panels of Figure \ref{fig:vdensity-fsourcesPDFs}
         we show the ratio of the volume occupied by cells with 
         a given $x_{\mathrm{HII}}$ over the total volume at $t$ = 100, 200 and 
         500~Myr. Note that the PDFs reflect the trend of 
         Figure \ref{fig:vdensity-fsources}, with the largest (smallest) volume occupied by ionised cells in the least (most) refined configuration.
         Also note that the 
         evolution in time of ionised regions reduces the differences while maintaining the trends across refinement levels. 
     
 \begin{figure} 
 \subfloat{\includegraphics[width=\columnwidth, height=16cm]{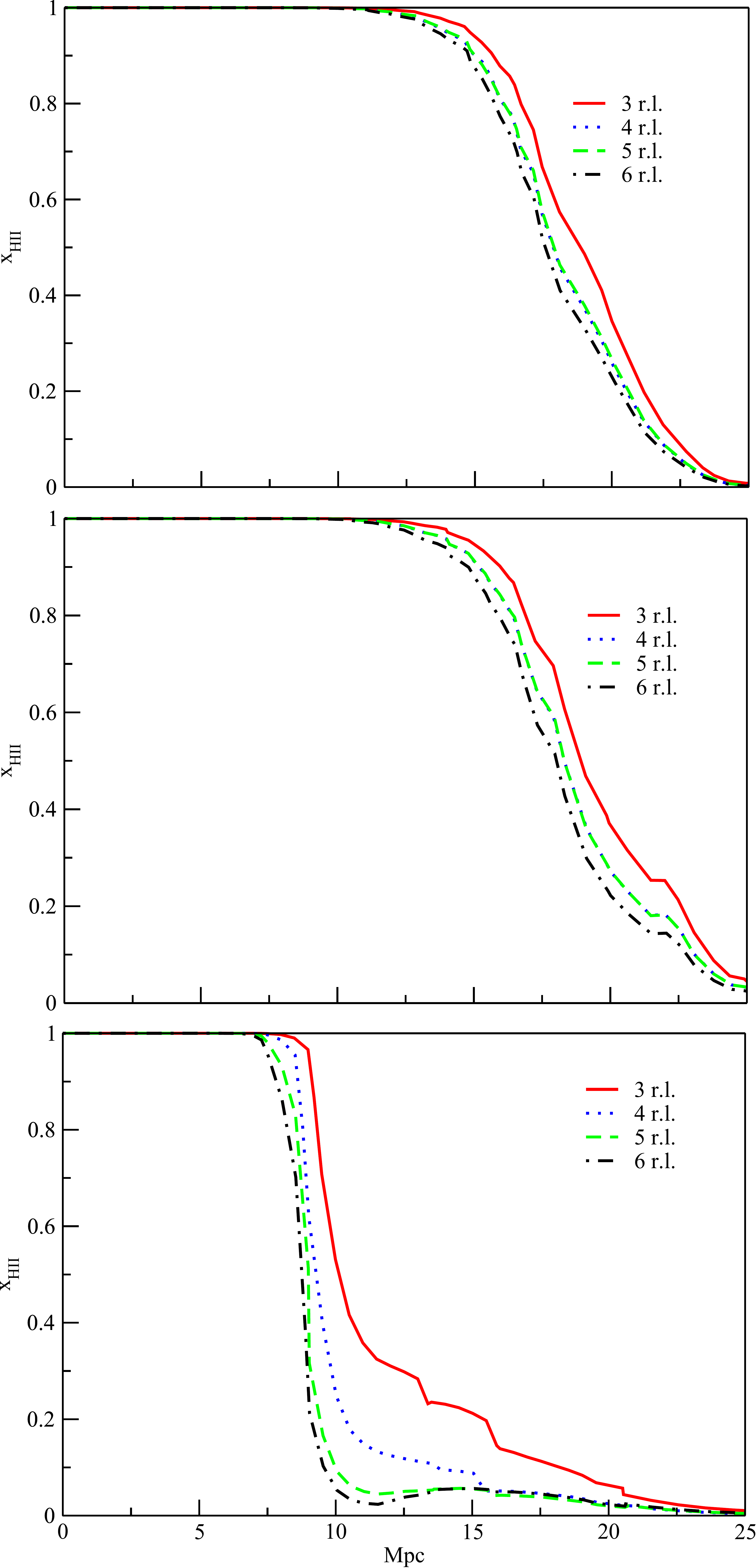}}
 \setlength{\abovecaptionskip}{10pt}
 \captionsetup{width=\columnwidth}
 \caption{$x_{\mathrm{HII}}$ profiles at time $t=500$~Myr for three LOSs in
random directions from the most luminous point source in Test~2a. The lines
refer to a test case with three
   (solid red), four (dotted blue), five (dashed green) and six (dot-dash black)
refinement levels.}
\label{fig:vdensity-fsources-los}
\end{figure}

Figure \ref{fig:vdensity-fsources-los} shows the $x_{\mathrm{HII}}$ profile for
three randomly selected line of sights (LOSs) from the most luminous point
source in the simulation. The trend observed in the LOSs reflects the above comments. In all the panels,
the ionisation front (I-front\footnote{We define the I-front as the 
point at which the $x_{\mathrm{HII}}$ drops below 0.5.})
systematically recedes with a larger number of refinement levels, due to the
increased gradient in $n_{\mathrm{gas}}$ and the higher gas recombination rate. 
We note that, although the above trend is observed in the vast majority of the
LOSs, there are exceptions due to the large variety of gas properties. We find
that the distance from the source location to the I-front position for the test cases with 6 
refinement levels is smaller than that with 3 refinement levels by 6.8, 4.39 and 13.6\% from the top
to the bottom panel.                   

We then conclude that the photo-ionisation algorithm of \texttt{CRASH-AMR} is
highly sensitive to the changes in the gas number density resolved by
more refinement levels; this is reflected in the variations observed in the
ionisation structures and I-fronts. \texttt{CRASH-AMR} will provide a more
precise and realistic representation of how the ionised bubbles form and grow
around the high density regions in which star formation occurs, as well as a
better estimate of the escape of ionising photons through the IGM when
local-scale reionisation simulations will be performed with this technique.

\begin{figure*}
 \vspace*{.1cm}
 \subfloat{\includegraphics[width=\textwidth, height=3.2cm]{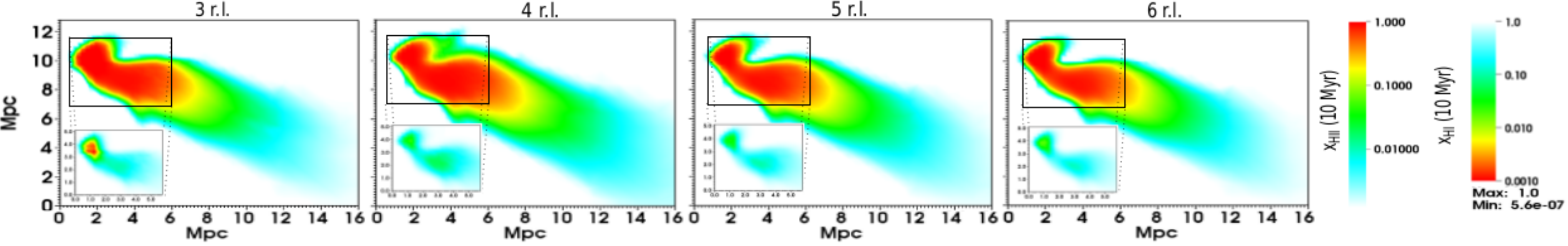}}
 \setlength{\abovecaptionskip}{10pt}
  \caption{Maps of $x_{\mathrm{HII}}$ for Test~2b at time $t=10$~Myr. From left
to right, the columns refer to simulations run with three, four, five and six
refinement levels (see text for more details). Panel insets show the neutral fraction of the gas in the zoom-in areas; 
their colour coding refers to the second palette on the right. Min and max values, 
when not represented by the palette, are written as reference.}
 \label{fig:vdensity-gsources}
\end{figure*}

\subsubsection{Test~2b: multiple sources set close to each other - constant $T$}
\label{subsubsection:vary-density-gather-sources}

Differently from Test~2a, here we place the twenty sources so that they are
close enough at the highest resolution to be gathered at lower refinement
levels. This results in 12, 6 and 2 sources at the 5th, 4th and 3rd refinement
level, respectively. Here we assume $\dot{N_\gamma}_\mathrm{ref}$ = $5 \cdot
10^{53} \mathrm{photons} \cdot \mathrm{s^{-1}}$, while the rest of the set-up is
the same as in Test~2a. 

Since in this test multiple sources are represented at the coarser levels as a
single one of higher luminosity, at early times we expect to see the growth of
only one ionised region at the coarser level, whereas at higher levels the
ionised regions remain distinct from each other. To capture these features, in
Figure \ref{fig:vdensity-gsources} we show maps of $x_{\mathrm{HII}}$ at
$t=10$~Myr. While at low resolution we find a single ionised region, with
$x_{\mathrm{HII}}$ in the range $\sim$0.8-1, at higher resolutions a much
smaller region has such high ionisation fraction. This translates into a volume
averaged $x_{\mathrm{HII}}$ fraction of 1.38, 1.43, 1.38 and 1.37 $\cdot
10^\mathrm{-3}$ for the 3rd, 4th, 5th and 6th refinement level, respectively, 
with a
0.7\% difference between the 3rd and 6th levels. At 500~Myr, the differences in
the $x_{\mathrm{HII}}$ fraction averages are 1.1\%.
Panel insets show the 
distribution of neutral gas in the zoomed regions. Note that here the colour coding 
shows areas with $x_{\mathrm{HI}} < 10^{-3}$ in red, while white areas represent 
almost neutral regions. Also note how the increased refinement progressively confines the 
ionised areas at the center.

We have compared the PDFs of $x_{\mathrm{HII}}$ between the different test cases and 
find a trend similar to the one of Test~2a, i.e., the configuration with 6 r.l. shows the least volume occupied by
fully ionised cells.  

Our comparison of the LOSs between the different test cases also shows a trend
similar to Test~2a, i.e., the largest differences are observed in the partially
ionised gas, with the extent of the ionised region receding with increasing
resolution.

\begin{figure*}
 \vspace*{.1cm}
 \subfloat{\includegraphics[width=\textwidth]{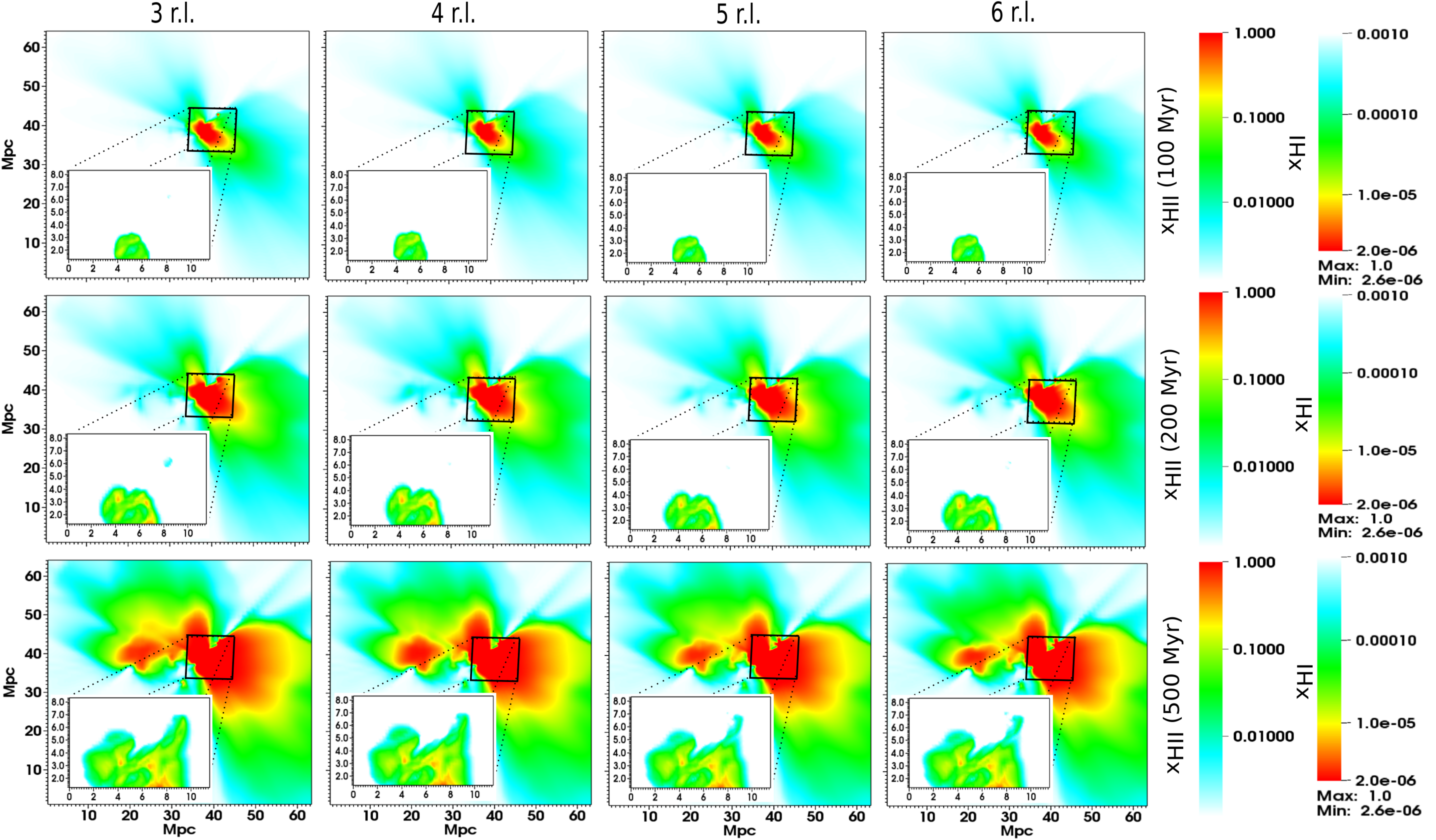}}
 \setlength{\abovecaptionskip}{10pt}
  \caption{Maps cut through the simulation volume for Test~2c. \textbf{Top:}
Maps of $x_{\mathrm{HII}}$ at time $t=100$~Myr. \textbf{Middle:} Maps of
$x_{\mathrm{HII}}$ at time $t=200$~Myr. \textbf{Bottom:} Maps of
$x_{\mathrm{HII}}$ at time $t=500$~Myr. From left to right, the columns refer to
simulations run with three, four, five and six refinement levels (see text for
more details). Panel insets show the neutral 
fraction of the gas in the zoom-in areas; their colour codings refer to the second 
palette on the right. Min and max values, when not represented by the palette, are 
written as reference.}
 \label{fig:vdensity-sources-xHII}
\end{figure*}

\begin{figure*}
 \vspace*{.1cm}
 \subfloat{\includegraphics[width=\textwidth]{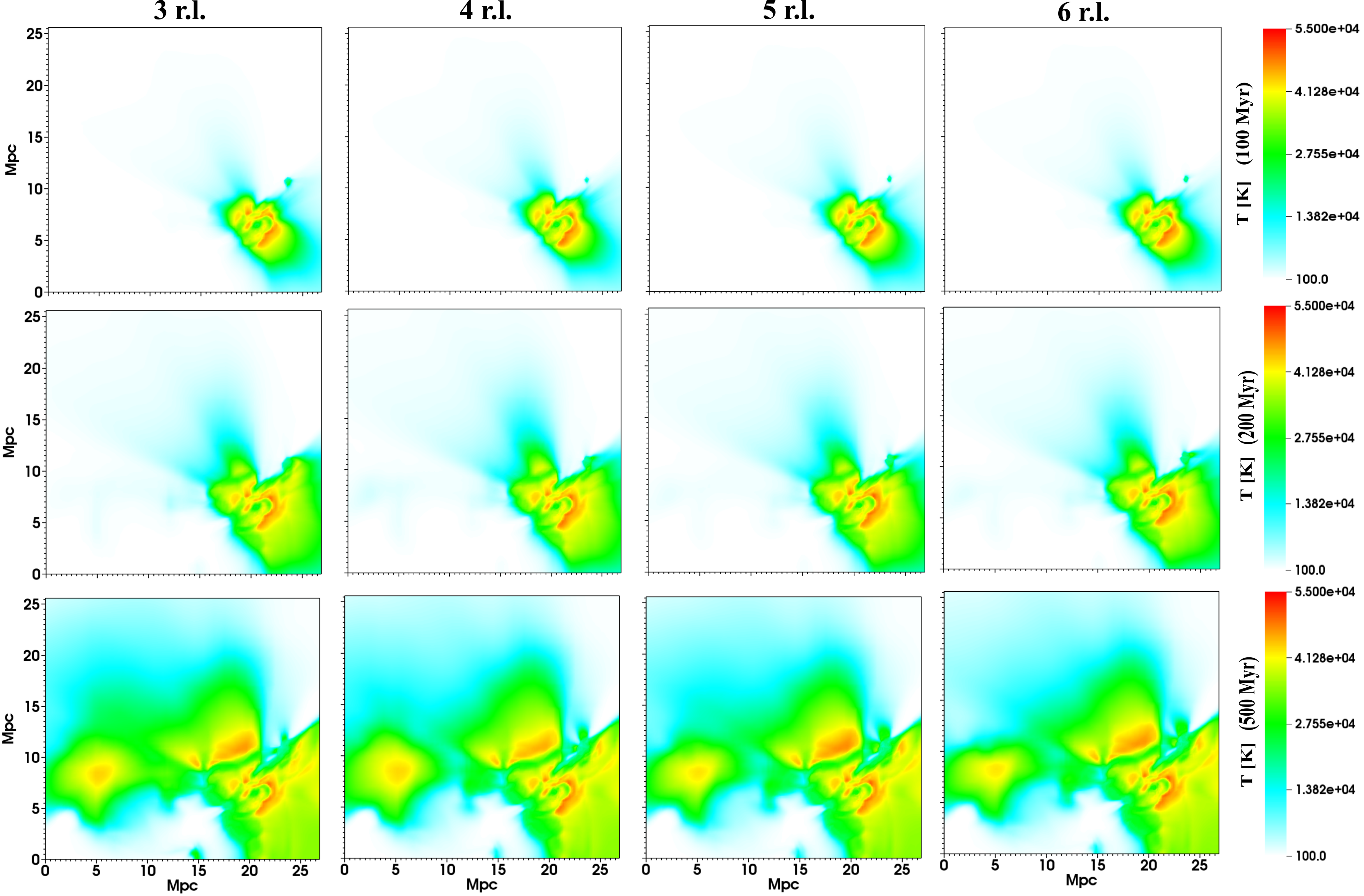}}
 \setlength{\abovecaptionskip}{10pt}  
 \caption{Maps cut through the simulation volume for Test~2c. \textbf{Top:} Maps
of $T$ at time $t=100$~Myr. \textbf{Middle:} Maps of $T$ at time $t=200$~Myr.
\textbf{Bottom:} Maps of $T$ at time $t=500$~Myr. From left to right, the
columns refer to simulations run with three, four, five and six refinement
levels (see text for more details).}
 \label{fig:vdensity-sources-Temp}
\end{figure*}

\subsubsection{Test~2c: multiple sources set far apart - $T$ calculated
self-consistently}
\label{subsubsection:vary-density-far-sources-Temp}

The sources here are located as in Test~2a, but now they have a black-body
spectrum at temperature $T_\mathrm{BB} = 10^\mathrm{5}$~K and the gas 
temperature is
calculated self-consistently with the progress of ionisation during the
simulation. 

 \begin{figure} 
 \subfloat{\includegraphics[width=\columnwidth, height=16cm]{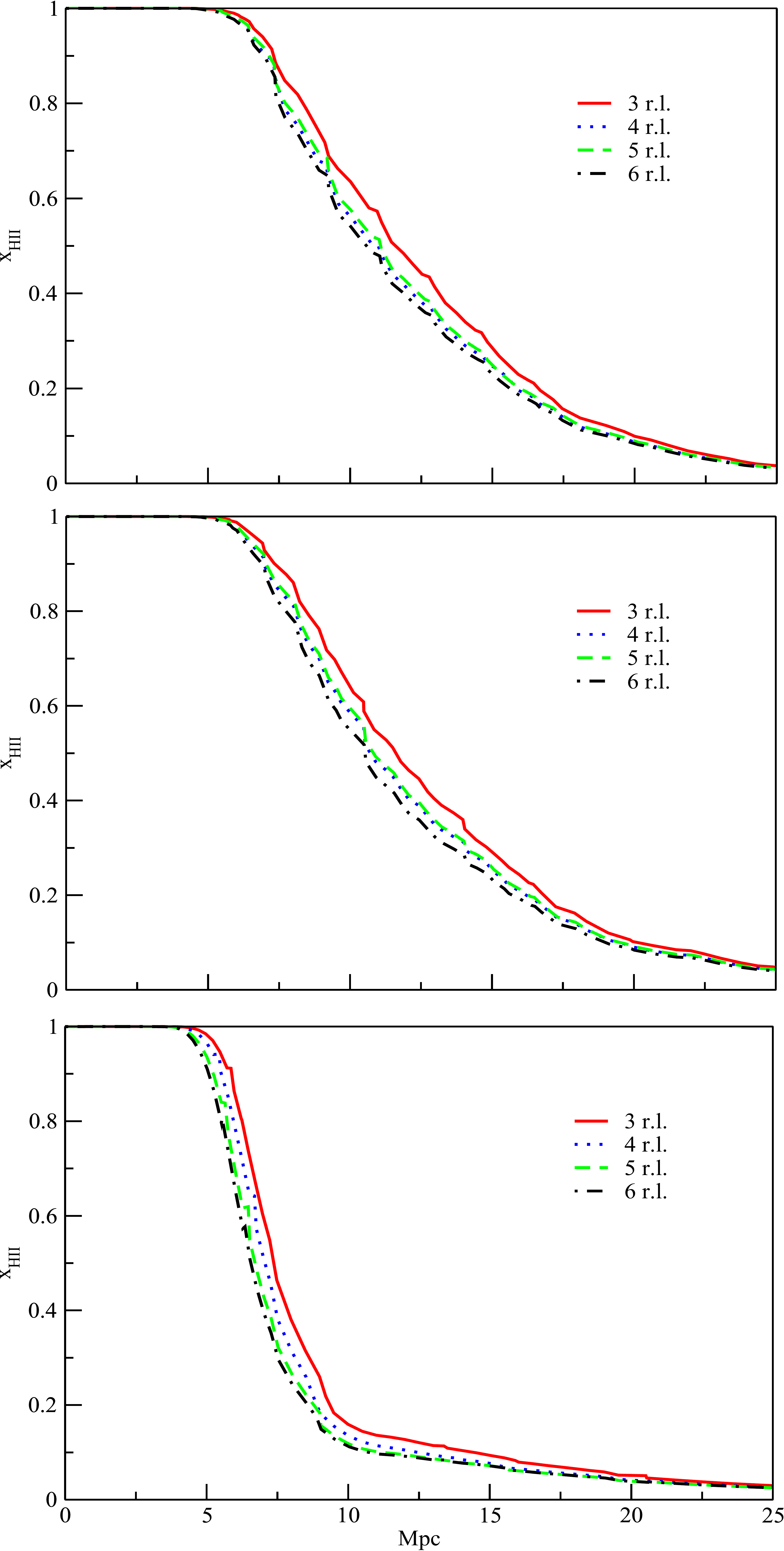}}
 \setlength{\abovecaptionskip}{10pt}
 \captionsetup{width=\columnwidth}
 \caption{$x_{\mathrm{HII}}$ profiles at time $t=500$~Myr for three LOSs in
random directions from the most luminous point source in Test~2c. The lines
refer to a test case with three
   (solid red), four (dotted blue), five (dashed green) and six (dot-dash black)
refinement levels.}
\label{fig:vdensity-fsources-temp-los}
\end{figure}

 \begin{figure} 
 \subfloat{\includegraphics[width=\columnwidth, height=16cm]{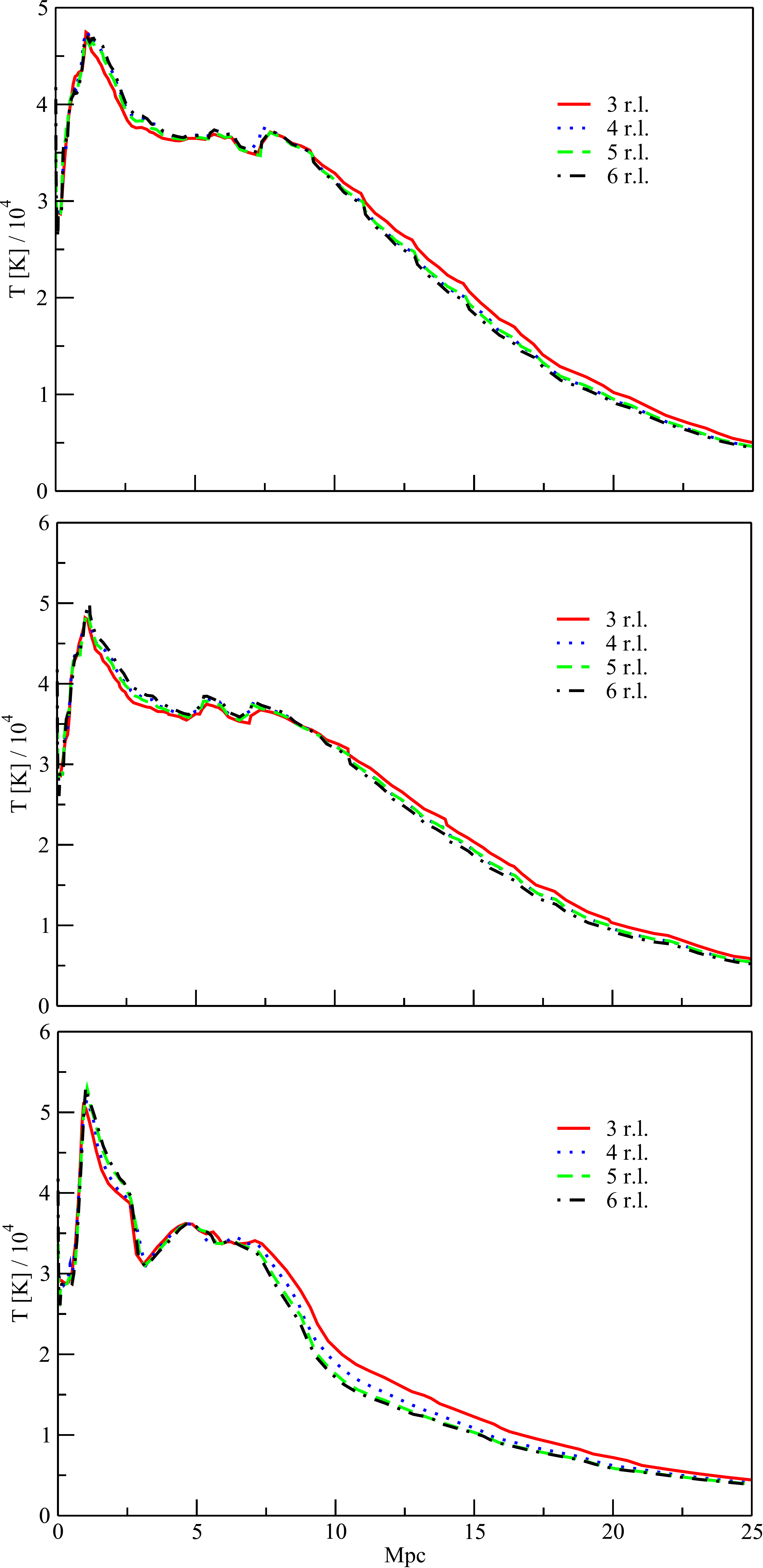}}
 \setlength{\abovecaptionskip}{10pt}
 \captionsetup{width=\columnwidth}
 \caption{$T$ profiles at time $t=500$~Myr for three LOSs in random directions
from the most luminous point source in Test~2c. The lines refer to a test case
with three
   (solid red), four (dotted blue), five (dashed green) and six (dot-dash black)
refinement levels.}
\label{fig:vdensity-fsources-tempA-los}
\end{figure}

Figures \ref{fig:vdensity-sources-xHII} and \ref{fig:vdensity-sources-Temp} show
the maps of $x_{\mathrm{HII}}$ and $T$ created in simulations with different
refinement levels at a time $t=100$, $200$~Myr and 500~Myr. We show the same
slices of Figure \ref{fig:vdensity-fsources}, i.e., containing the most
luminuous point source. Hence, the density field corresponds to the bottom slice
in Figure \ref{fig:vdensity-fsources}.

Similarly to Test~2a, we find that at a time $t=100$~Myr the maps look very
similar. At $t=200$~Myr, the ionised regions start to show some differences,
noticeably in the separate bubble that is formed in all the slices. The same
differences can be found in the maps of $T$ as well. At $t=500$~Myr, the most
obvious differences are found in the H\(\ions{II} \) region on the left that is
formed by a point source lying on a plane different from the one shown in the
Figure. The corresponding $T$ maps have a similar behaviour. Finally note 
that in each panel we provided a zoom-in view of the central region showing the 
residual distribution of neutral gas. 
As discussed above, many areas with residual neutral gas are found
in the inner region, with values in the range $10^{-5} < x_{\mathrm{HI}} < 10^{-4}$.

The differences are reflected also in the volume averaged ionisation fraction,
which is $x_{\mathrm{HII}}$=2.54, 2.49, 2.39 and 2.34 $\cdot 10^\mathrm{-2}$ for 
3, 4,
5 and 6 refinement levels at time $t=500$~Myr, respectively, with a 8.5\%
difference between the 3rd and 6th levels. The corresponding temperatures are
2264, 2221, 2189 and 2159 K, with a 4.8\% difference between the 3rd and 6th
levels. 
Comparing Figures \ref{fig:vdensity-fsources} and
\ref{fig:vdensity-sources-xHII} we find that the size of the H\(\ions{II} \)
region is smaller in the latter case, as a result of the spectral distribution
of the ionising sources and the self-consistent calculation of the gas
temperature: the introduction of a spectral distribution creates a number of
high-energy photon packets which can easily diffuse beyond the ionisation fronts
creating wider regions at low ionisation, while the various cooling processes
which are being taken into account have a substantial feedback on the gas
recombination rates creating smoother ionisation gradients. 
We do expect these differences to be even more marked in the presence of Helium.

In Figure \ref{fig:vdensity-fsources-temp-los} we show $x_{\mathrm{HII}}$ along
the three LOSs of Figure \ref{fig:vdensity-sources-xHII}. The corresponding $T$
is shown in Figure \ref{fig:vdensity-fsources-tempA-los}. A trend similar to the
one of Test~2a is observed, i.e., the extent of the fully ionised region becomes
smaller with increasing refinements levels. The $T$ profiles exhibit the same
behaviour, with a good agreement within the fully ionised region and a $T$
decreasing with increasing refinement levels.
As already mentioned in the description of the test setups, the base grid 
     in use is refined only in the central region, close to the sources. As a 
     consequence, the resolution at large distances is the same in all cases, 
     and the discrepancies are due to the differences experienced by the various
     photon packets during their propagation through the central region rather than to local density 
     variations.
     
\section{Conclusion}
\label{section:Conclusion}

In this paper we have introduced \texttt{CRASH-AMR}, a new version of the
cosmological radiative transfer code \texttt{CRASH}, enabled to run RT
simulations on AMR grids. After an exhaustive discussion of the code, we have
shown the results of many tests both with the simplified set-up prescribed in
the Radiative Transfer Code Comparison Project and a realistic hydrodynamic
simulation with AMR refinement. All the tests show good agreement with the
latest release of \texttt{CRASH}, confirming the correct inclusion of a more
accurate and alternative geometry representation of the gas distribution in the
cosmological domain in which the RT simulation is performed. The small
discrepancies found are due either to the presence of a grid at higher
resolution in the refined levels or to averaging operations. 

The application to a realistic density field shows differences in the pattern of
the ionised regions because of the more accurate treatment of the gas optical
depth and cooling function. Consequently, at higher resolution the gas
ionisation fractions and temperature are calculated with greater accuracy,
allowing a better modeling of the growth of ionised bubbles, as well as of the
escape of ionising radiation from high-density regions in which star formation
is typically embedded. In general, \texttt{CRASH-AMR} will provide an invaluable
improvement compared to the previous algorithm whenever a better resolution of
the radiation-matter interaction in some specific regions is needed.

As final consideration, we note that \texttt{CRASH-AMR} is able to perform RT
simulations in high density regions with the resolution increased by a factor of
64 with respect to the base $64^\mathrm{3}$ grid resolution without experiencing
serious memory limitations. Such a high resolution would be unmanageable from
the storage point of view in a single CPU core without this new version of the
code. We also find that by running a specific configuration on smarter AMR grids
rather than on uniform $512^\mathrm{3}$ grids, a $\sim$60\% reduction in run 
time is
obtained. Hence, \texttt{CRASH-AMR} provides an advantage both in terms of
memory consumption and run time performance when compared to the standard
version of \texttt{CRASH}, especially in the future releases where
\texttt{CRASH-AMR} will take advantage of distributed memory parallelism using
MPI.  

\section*{Acknowledgements}

N.H thanks the \texttt{CHOMBO} and \texttt{FLASH} development team for their
help in using \texttt{CHOMBO} with a Fortran code. N.H also thanks Koki Kakiichi
and Prof. Michael Bader for their useful suggestions. LG acknowledges the support 
of the DFG Priority Program 1573 and the support of the European Research Council 
under the European Union (FP/2007-2013) / ERC Grant Agreement n. 306476.

{
\bibliography{amrpaper}
}

\appendix
\section{Dependence on grid resolution}
\label{section:Grid-resolution}

The difference in results observed in Section~\ref{subsection:Test1-AMR} is only
due to the grid resolution and not to the new \texttt{CRASH-AMR} implementation.
While some discrepances are expected when the tests are set-up with different
base grid resolutions, this should not be the case when the resolution at the
finest refinement level is the same. 
We demonstrate this by setting up some test cases similar to Test 1, where the
resolution at the base grid is different, but that at the finest AMR level is
the same, i.e. 512$^\mathrm{3}$. We thus expect the RT simulations to give the 
same
results. Note that here we have completely refined the grid, although this is
not usually done with AMR codes. We set up test cases with the following grid
properties:

\begin{enumerate} [(a)]
 \item base grid resolution $64^\mathrm{3}$, three levels of refinement;
 \item base grid resolution $128^\mathrm{3}$, two levels of refinement;
 \item base grid resolution $256^\mathrm{3}$, one refinement level;
 \item base grid resolution $512^\mathrm{3}$, no refinement.
\end{enumerate}

\begin{figure} 
 \subfloat{\includegraphics[width=0.98\columnwidth]{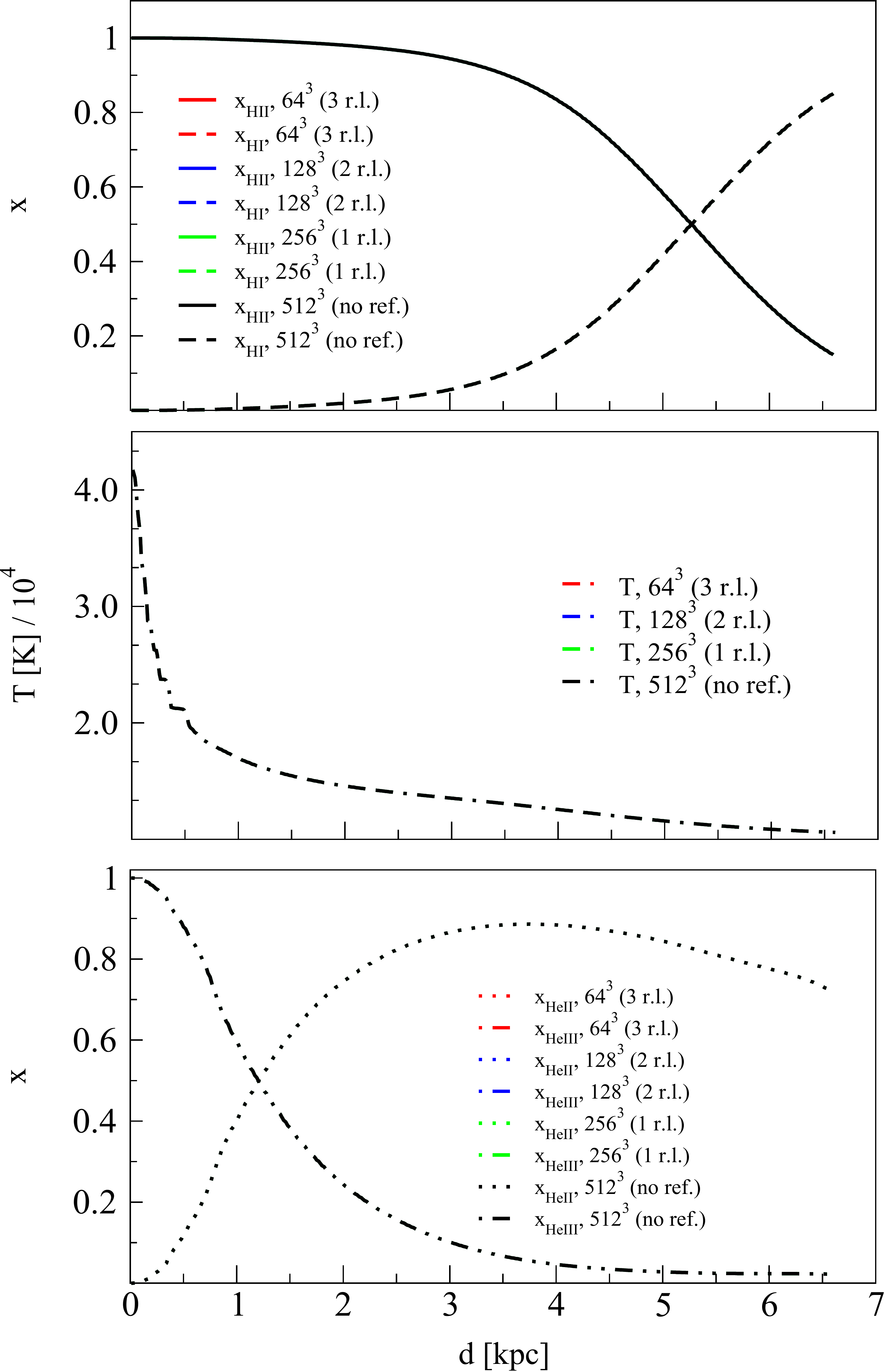}}
 \setlength{\abovecaptionskip}{10pt}
 \caption{Spherically-averaged profiles at time $t=500$~Myr for Test~1b. The
lines refer to \texttt{CRASH-AMR} with different base grid resolutions, 
$64^\mathrm{3}$
(four AMR levels - red), $128^\mathrm{3}$ (three AMR levels - blue), 
$256^\mathrm{3}$ (two AMR
levels - green) and $512^\mathrm{3}$ (no refinement - black). \textbf{Top:} 
Profiles of
$x_{\mathrm{HII}}$ (solid lines) and $x_{\mathrm{HI}}$ (dashed lines).
\textbf{Middle:} Profile of $T$ (dash-dot-dash lines). \textbf{Bottom:} Profiles
of $x_{\mathrm{HeII}}$ (dotted lines) and $x_{\mathrm{HeIII}}$ (dot-dash
lines).}
\label{fig:sphericalT2-sameAMR512}
\end{figure}

Figure~\ref{fig:sphericalT2-sameAMR512} shows the spherical averages  of
$x_{\mathrm{HII}}$, $x_{\mathrm{HI}}$, $T$, $x_{\mathrm{HeII}}$ and
$x_{\mathrm{HeIII}}$ from the above simulations. We have calculated the
spherical average at the highest resolution for all cases, and, as expected, it
is exactly the same for the different base grid resolutions. 

\section{Code performance}
\label{run-time}

\begin{table*}
    \caption{Run times (in hours) and the corresponding volume averaged
ionisation fraction $x_{\mathrm{HII}}$ at time $t=500$~Myr for
\texttt{CRASH-AMR} simulations run on a configuration similar to that in
Test~2a.}
    \centering
    \begin{tabular}{ | c | p{3.5cm} | p{2.5cm} | p{2.5cm} | p{2.5cm} }
    \hline
    Test case & Grid resolution & Run time (hours) & $x_{\mathrm{HII}}$ \\ 
    \hline
    \textit{U128} & Uniform grid (UG), $128^\mathrm{3}$ & 1.60 & 0.181 \\ 
    \textit{R128} & Refined grid (RG), $128^\mathrm{3}$ & 1.58 & 0.180 \\ 
    \textit{U256} & UG, $256^\mathrm{3}$ & 3.55 & 0.185 \\ 
    \textit{R256} & RG, $256^\mathrm{3}$ & 2.52 & 0.184 \\ 
    \textit{U512} & UG, $512^\mathrm{3}$ & 8.32 & 0.183 \\ 
    \textit{R512} & RG, $512^\mathrm{3}$ & 3.40 & 0.181 \\ 
    \hline
    \end{tabular}
     \label{tbl:Table-runtime}	
\end{table*}

In this section we investigate the run time performance of \texttt{CRASH-AMR}
while ensuring correctness in results. In Section~\ref{section:CRASH-CHOMBO} we
mentioned that there was no run time overhead from the coupling of
\texttt{CHOMBO} to \texttt{CRASH}, but only the additional time to search for
the new cell in the multiple refinement levels that the ray crosses. Even though
this search adds to the total run time in the RT simulations done using AMR
grids, its effect is small when compared to running the RT simulation on a
uniform high resolution grid. To prove that this is indeed the case, we ran some
tests comparing the run times between grids at uniform resolution and AMR grids
from the \texttt{CHARM} simulation. We then compare the two test cases to ensure
that \texttt{CRASH-AMR} is able to provide accurate results.

\subsection{Set up with a single point source}

The set-up is similar to Test~2a, with the ionised region from a single point
source expanding into a realistic density field. The source is placed in the
highest density peak, with \textit{$\dot{N_\gamma}$} = $5 \cdot 10^\mathrm{54}
\mathrm{photons}\cdot \mathrm{s^{-1}}$, a monochromatic spectrum of $E_{\nu}$ =
13.6~eV and emits $N_\mathrm{p} = 2 \cdot 10^\mathrm{8}$ photon packets. The gas 
is assumed to
contain ony H with an initial ionisation fraction $x_{\mathrm{HII}}$ set to $1.2
\cdot 10^\mathrm{-3}$. The gas temperature $T$ is initially set to 100~K and is 
kept
constant throughout the simulation. The simulation time is 
$t_\mathrm{sim}=500$~Myr,
starting at redshift $z$ = 0.1. We output the results at intermediate times $t$
= 10, 20, 100, 200 and 500~Myr.

To compare the run times we have created uniform resolution grids from the AMR
refined grids used in Test~2 by interpolating the coarse data onto the finer
levels. We use uniform grids of resolution $128^\mathrm{3}$ (\textit{U128}), 
$256^\mathrm{3}$
(\textit{U256}) and $512^\mathrm{3}$ (\textit{U512}), and compare the run times 
to
those of grids with base resolution of $64^\mathrm{3}$ and 1 (\textit{R128}), 2
(\textit{R256}) and 3 (\textit{R512}) refinement levels, respectively. We would
like to point out that there are other factors that also impact the run time of
a \texttt{CRASH} simulation, for example the number of point sources, presence
of He with $T$ evolution and number of photon packets. However, we do not
consider them here and focus instead on the performance of the code for a simple
test case.
Table \ref{tbl:Table-runtime} shows the resulting run times together with the
corresponding volume averaged ionisation fraction $x_{\mathrm{HII}}$ at time
$t=500$~Myr. 

\begin{figure*}
  \centering
  \subfloat{\includegraphics[width=\textwidth, height=14cm]{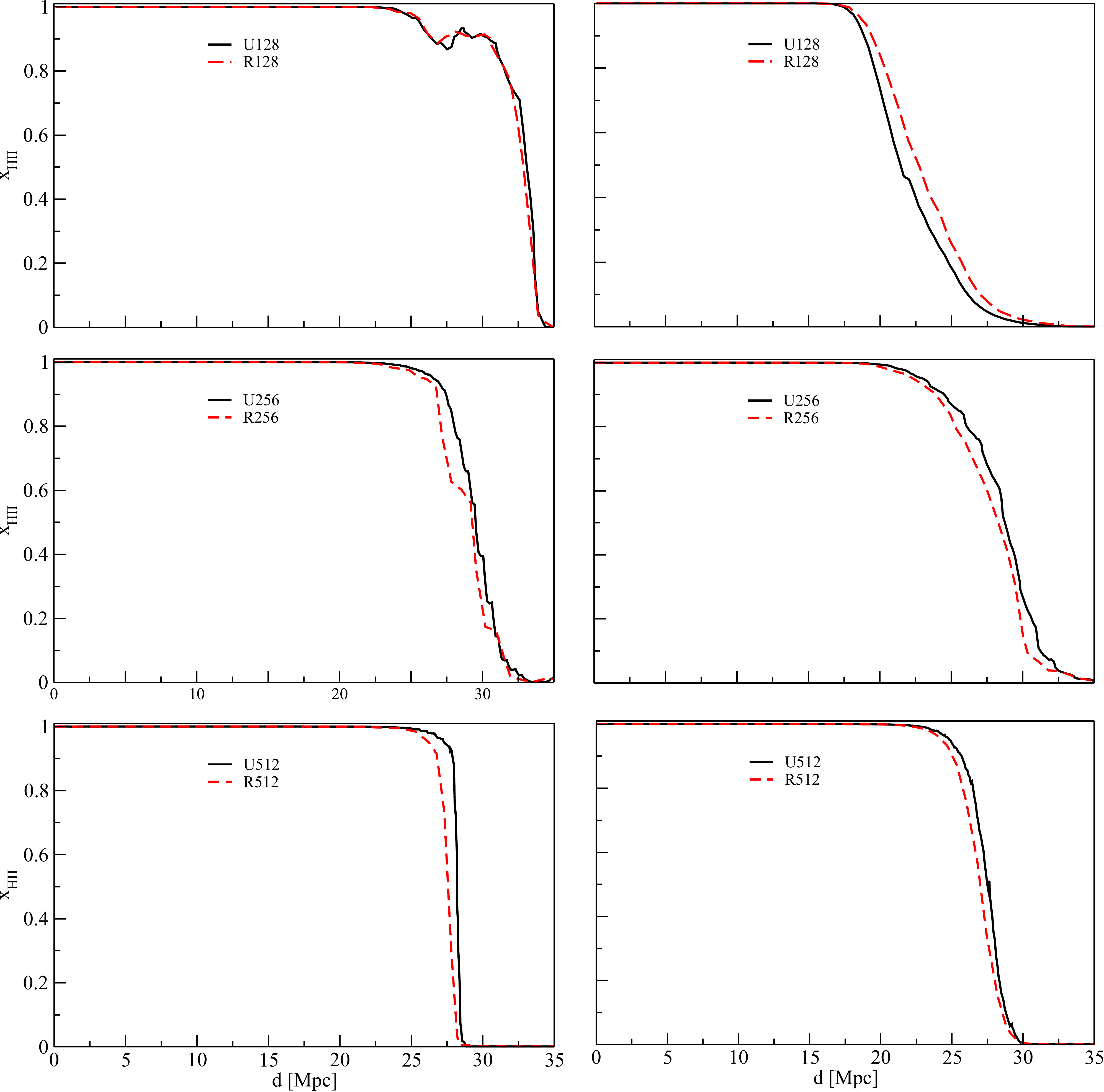}}
 \caption{LOSs in two random directions from a point source for Test~2a at time
$t=500$~Myr. The lines refer to \texttt{CRASH-AMR} with a uniform grid (solid
black) and with refinement levels (dashed red). \textbf{Top:} \textit{U128} and
\textit{R128}. \textbf{Middle:} \textit{U256} and \textit{R256}.
\textbf{Bottom:} \textit{U512} and \textit{R512}.}
\label{fig:compare-los-single}
\end{figure*}

We find that the difference in run time between \textit{U128} and \textit{R128}
is only 1.25\%, but this increases as we move to higher resolutions, and reaches
59\% for \textit{U512} and \textit{R512}.
This large gain in computational speed does not come at the expense of
correctness of results. In fact, the difference between the uniform and refined
grids in terms of volume average ionisation fraction is only 0.5, 0.5 and 1.1\%.
Additionally, two random LOSs are compared in Figure
\ref{fig:compare-los-single} for the three different resolutions. We find that
the extent of the fully ionised H\(\ions{II} \) region for the uniform and AMR
grid cases is the same at all resolutions, while the partially ionised regions
show some discrepancy. The I-front in the refined grids is in fact smaller than
that in the uniform grids, with a maximum difference of 2\%.  

\subsection{Set up with multiple point sources}

Here we use again a setting similar to that of Test~2a, and compared the results
between a $512^\mathrm{3}$ uniform grid and a grid with a base resolution of 
$64^\mathrm{3}$
with 3 refinement levels.
Table \ref{tbl:Table-runtime-multiple-ps} shows the run times together with the
corresponding $x_{\mathrm{HII}}$ at time $t=500$~Myr.

\begin{figure}
  \centering
   \subfloat{\includegraphics[width=0.98\columnwidth]{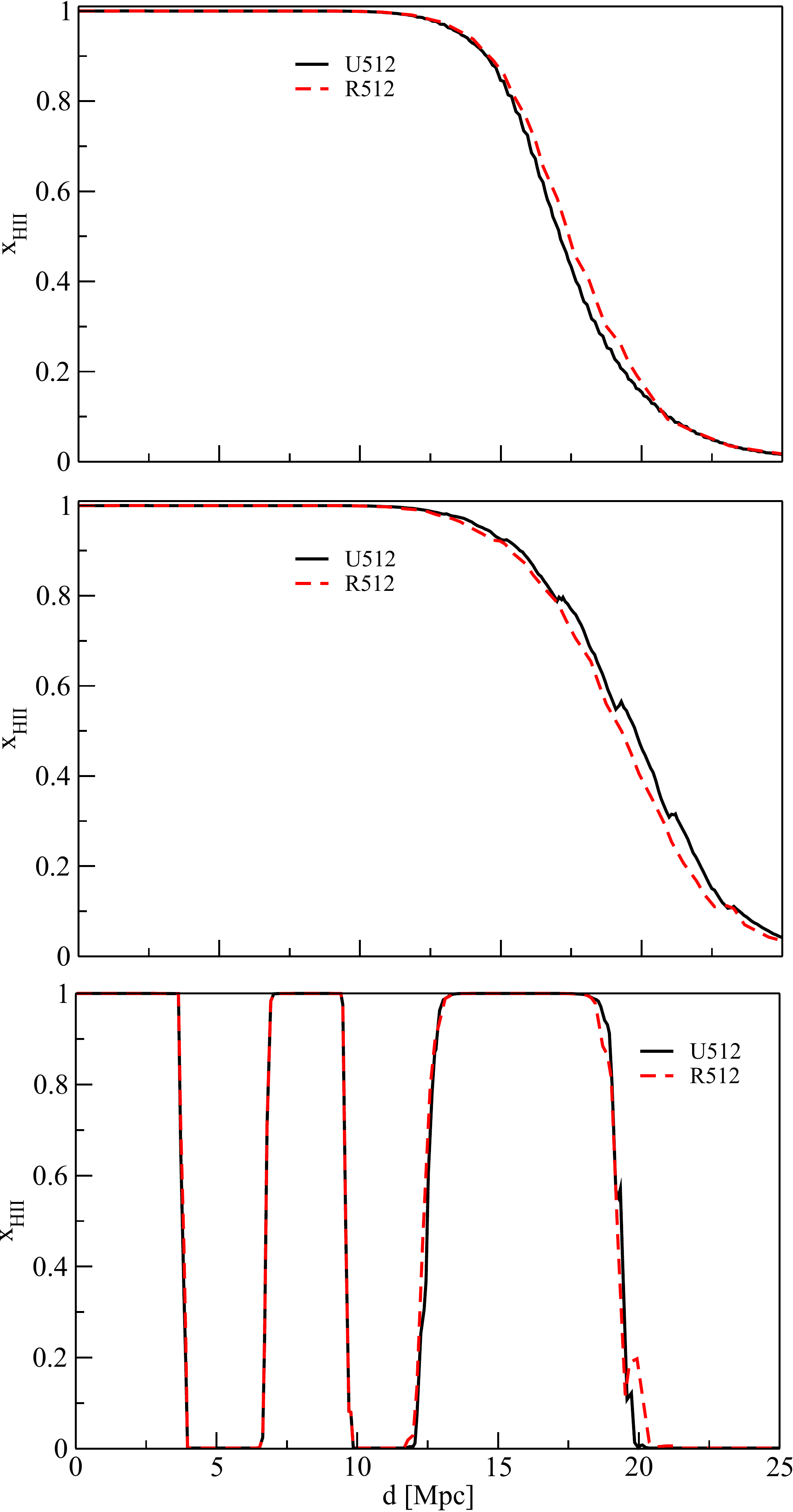}}
 \caption{LOSs in random directions from a point source for Test~2a at time
$t=500$~Myr. The lines refer to \texttt{CRASH-AMR} with a uniform grid (solid
black) and with 3 r.l. (dashed red).}
\label{fig:compare-los-multiple}
\end{figure}

Here again we find that the difference in run times and volume average
ionisation fraction between the two cases is as high as 38\% and 1.3\%,
respectively. A comparison between three random LOSs from the most luminous
source (Fig. \ref{fig:compare-los-multiple}) confirms that the extent of the
fully ionised regions is the same with and without uniform grids, while the
maximum difference in the size of the I-front is 2\%. 

\begin{table*}
    \caption{Run times (in hours) and the corresponding $x_{\mathrm{HII}}$ at
time $t=500$~Myr for \texttt{CRASH-AMR} simulations run on a configuration
similar to that in Test~2a.}
    \vspace{0.5cm}
    \centering
    {\renewcommand{\arraystretch}{1.2}
    \begin{tabular}{ | c | c | c | c | c }
    \hline
    Test case & Grid resolution & Run time (hours) & $x_{\mathrm{HII}}$ \\ 
    \hline
    \textit{U512} & UG, $512^\mathrm{3}$ & 11.4 & 0.0501 \\ 
    \textit{R512} & RG, $512^\mathrm{3}$ & 6.97 & 0.0494 \\ 
    \hline
    \end{tabular}
    }
     \label{tbl:Table-runtime-multiple-ps}	
\end{table*}

From the two test cases above it is clear that \texttt{CRASH-AMR} provides
results consistent with those from \texttt{CRASH3} but with much shorter run
times. We have been able to represent the regions of interest at a resolution
increased by a factor of 64 and not lose any accuracy in the
results\footnote{Note here that the accuracy of the results is very problem
dependent.}. Our interface with \texttt{CHOMBO} and the ray-tracing algorithm to
search for the new cell across different refinement levels during ray-tracing is
highly efficient and does not have a negative impact on the run time.

We would, however, like to point out that in both the test cases above the decrease in run times 
between the uniform resolution and AMR grids is not proportional to the percentage decrease in 
the number of cells in the AMR grids. For example, in test case \textit{U512} the number of 
cells in the grid is 134,217,728 while that in \textit{R512} is 1,816,576 which is a 98\% decrease
in the number of cells. However, the corresponding decrease in run times is 60\%. This is due to the fact that the
run time now includes:

\begin{enumerate} [(a)]
  \item the time taken for the initial setup, which includes setting up the AMR hierarchy in \texttt{CRASH-AMR}. 
                This is done once at the beginning of the simulation;
  \item \label{search} the time taken to find the right box that the ray is in, every time it enters a new cell. This step includes a number
        of other checks, which have been discussed in section \ref{section:CRASH-CHOMBO}.       
\end{enumerate}

Also, as of now, we have implemented a simple search routine in \ref{search} to loop  
through an unsorted neighbor and parent lists until the correct box is found. As the PBAMR
scheme allows for multiple neighbor and parent boxes, this search can be expensive
if done frequently. We are looking at optimising this search by sorting the neighbor and
parent lists in the $3D$ space, by using the Morton space-filling curve implemented in \texttt{CHOMBO}, 
before using them to search for the new box.

\bsp
\label{lastpage}
\end{document}